%% file: main.tex
\documentclass{article}
\usepackage{authblk}

\input{shortcuts}

\title{Cross-Study Replicability in Cluster Analysis}
\author[*]{Lorenzo Masoero}
\author[1]{Emma Thomas}
\author[2]{Giovanni Parmigiani}
\author[2]{Svitlana Tyekucheva}
\author[2]{Lorenzo Trippa}
\affil[*]{Email address: \emph{lo.masoero@gmail.com}}
\affil[1]{Department of Biostatistics, Harvard T.H.\ Chan School of Public Health}
\affil[2]{Department of Data Science, DFCI}
\date{September 10, 2022}
\begin{document}
\maketitle
\input{body}
\bibliographystyle{abbrvnat}
\bibliography{ref_cl.bib}
\newpage
\appendix
\input{sec-add}

\input{sec-other_methods}

\end{document}

%% file: body.tex
\input{sec-intro}
\input{sec-theory_single}
\input{sec-theory_multi}
\input{sec-exp-synth}
\input{sec-exp-real}
\input{sec-disc}
\input{sec-acks}

%% file: sec-intro.tex
% !TEX root = main.tex

%======================================
\section{Introduction}
\label{sec:intro}
%======================================

Clustering, the task of partitioning data into distinct classes, is fundamental in a variety of fields and applications.  For example, in genomics, clustering procedures are used for exploratory analyses, dimensionality reduction and to identify interpretable groups within  high-dimensional  data, such as gene expression studies.

\par One of the difficulties in clustering, common to other techniques in unsupervised learning, is the ambiguity of the notion of success. In contrast to supervised learning, where ground truth measurements can be used to validate the performance of a learning procedure (e.g., the precision of a classifier), in unsupervised learning a direct measure of success is not available. In applications it is however crucial to identify criteria to assess the reliability of these unsupervised learning methods.

\par In this paper we examine the problem of quantifying the quality of cluster analyses through the lens of replicability. We consider as a motivating example clustering in gene expression studies aiming to identify cancer subtypes. In this context, a dataset is a high dimensional collection of gene expression profiles of different patients, and clustering analyses try to identify biologically relevant groups of observations. An ideal cluster analysis identifies cancer subtypes and in turn allows scientists to develop specific and effective treatment strategies for the different subtypes.

\par We introduce notation and relevant background in \Cref{sec:review}, where we provide a thorough review of the existing literature on clustering replicability. Next, in \Cref{sec:theory_multi}, we focus on the assessment of clustering replicability when multiple sources of data are available --- a question of increasing importance in biosciences, where collections of  datasets generated by different  research groups  and  institutions are  often available \citep{hayes2006gene, bernau2014cross, trippa2015bayesian, national2019reproducibility}.  We develop a novel method that can help understand whether cluster procedures ran across multiple datasets are replicable.  The idea  underlying the replicability metrics we employ is that clusters consistently identified by independent analyses of distinct datasets can be used as a criterion to assess  the replicability of the analysis. We show how to generate replicability summaries, representative  of  the  similarity  of  the  groups identified  by a clustering method across independent  analyses   of  the available  datasets. Our procedure  evaluates the replicability,  without  constraints  on the choice  of  the  clustering method, and across any number of datasets, at  both a \emph{global} scale, that is for the whole data collection, as well as at a \emph{local} scale, that is for an individual cluster. We test our method on synthetic data in \Cref{sec:exp-synth}, and we present an application in cancer research in \Cref{sec:exp-real}. We provide additional experiments, also using competing replicability methods in the Appendix. Code to replicate all our experiments can be found at {\url{https://github.com/lorenzomasoero/clustering_replicability}}.

%% file: sec-theory_single.tex
% !TEX root = main.tex

%======================================
\section{Clustering Replicability in a Single Study}
\label{sec:review}
%======================================

With the growth of high dimensional and multi-modal datasets in several areas of science, practitioners need replicable procedures to analyze and simplify their data. In biological sciences, for example, advances in data-collection technologies allow investigators to study increasingly complex datasets, with large sample sizes and feature lists (gene expression profiles, demographics, imaging). While these rich datasets come with the promise of providing new insights, they also present practical challenges. Analysts often need dimensionality reduction techniques to visualize and explore the data. In this context, clustering algorithms have emerged as a preeminent technique because of their scalability, ease-of-use and wide applicability. While several clustering algorithms are available, assessing the usefulness and quality of the results they produce is a difficult problem, which has received considerable interest in the literature. In this section, we focus on replicability of cluster analyses and review important recent contributions in the literature on this problem. 

%======================================
\subsection{Preliminaries and notation} \label{sec:notation}
%======================================

\paragraph*{Data} A dataset  $\Xtr = \{x_1,\ldots,x_n\}$ is an unordered collection of datapoints. An individual observation or datapoint $x_i = [x_{i,1},\ldots,x_{i,p}]$ is a $p$-dimensional vector, e.g., the gene expression profile of patient $i$, where $x_{i,r}$ is the expression level of the $r$-th gene of interest. 

\paragraph*{Clustering Algorithms} A clustering algorithm $\mathcal{A}$ is a procedure that takes as input a dataset $\Xtr$ and outputs a ``learned'' clustering function $\psi(\cdot; \AA,\Xtr): \R^p \to [k]:=\{1,2,\dots, k\}$. Notice that $\psi(\cdot; \AA,\Xtr)$ maps \emph{any} point $y \in \R^p$ to a class $\ell \in [k]$. 

\paragraph*{Partitioning via Clustering Functions} Recall that subsets $U_1,\ldots,U_k$ s.t.\ $U_j \cap U_\ell = \emptyset$ for $j \neq \ell$ and $\cup_{j=1}^k U_j = \Xte$ form a partition of $\Xte$. Given clustering algorithm $\AA$ and training dataset $\Xtr$, a partition of \emph{any} collection of data points $\Xte$ can be directly obtained by applying the learned clustering function $\psi(\cdot; \AA,\Xtr)$ to all points $x' \in \Xte$. Datapoints in $\Xte$ sharing the same cluster label belong to the same subset of the partition, e.g.\ the $j$-th subset is given by $U_j = \{x' \in \Xte:\psi(x';\AA,\Xtr) = j\}$. We let $\Psi(\Xte; {\AA,  \Xtr}):=\{\psi(x'; \AA,  \Xtr), x' \in \Xte\}$ denote the labels of datapoints in $\Xte$ induced by the clustering function $\psi(\cdot; \AA,\Xtr)$.  In what follows, we write $\psi(\cdot)$ or $\psi(\cdot;\Xtr)$ in place of $\psi(\cdot; \AA,\Xtr)$ and $\Psi(\cdot)$ or $\Psi(\cdot;\Xtr)$ in place of $\Psi(\cdot; \AA,\Xtr)$ when $\AA$, $\Xtr$ are clear from the context. 

\paragraph*{Binary Partitions}  For $y, w \in \R^p$ and a clustering function $\psi(\cdot; \AA, \Xtr)$, we define the co-clustering operator $\tilde{\psi}_{w}(y;  \AA, \Xtr)= \ind \left\{\psi(y;  \AA, \Xtr) = \psi(w;  \AA, \Xtr) \right\}$.

\paragraph*{Example: $k$-means}  Let $\AA$ be the $k$-means algorithm. Given a training dataset $\Xtr$, this algorithm works by approximately solving the optimization problem:
\begin{align} \label{eq:opt_k_means}
	\min_{U_1,\dots,U_k} \sum_{\ell=1}^k \sum_{i \in U_\ell} \norm{x_i - {c}_\ell}_2,
\end{align}
where  $U_1,\dots, U_k$ are the disjoint subsets forming a partition of $\Xtr$, and for $\ell = 1,\ldots,k$, ${c}_\ell ~=~ [c_{\ell,1}, \ldots, c_{\ell,p}]$ is the $\ell$-th ``centroid'' of $U_\ell$, with components $c_{\ell,r} =\left(\sum_{i \in U_\ell} x_{i,r}\right)/|U_{\ell}| \in \R$, $r \in [p]$. Here, we view  $k$-means as the procedure that learns the clustering function 
\begin{align} \label{eq:predict}
	\psi(x'; \AA,\Xtr)=\argmin_{\ell \in \{1,\dots,k\}} \norm{x' - {c}_\ell}_2.
\end{align}

\paragraph*{Remark: Induced Clustering Functions} In some cases, the output of a clustering algorithm is not a clustering function, rather simply a partition of the training set (e.g.\ as in hierarchical clustering). In these cases we obtain a clustering function indirectly,  via an additional classification step. In the present work, for this class of algorithms, we adopt a nearest neighbor approach (although other classifiers could be used): let $z(i) \in [k]$ be the cluster label for datapoint $x_i$, $i=1,\ldots,n$ obtained by applying $\AA$ to $\Xtr$. Then, for a generic $y \in \R^p$, and $d(\cdot,\cdot)$ a distance metric on $\R^p$, we let $\psi(y) := z\left(\arg\min_{x_i \in \Xtr} d(y, x_i) \right)$.

%======================================
\subsection{Clustering Replicability via Stability} \label{sec:stability}
%======================================

Having introduced the necessary notation, we now start out review existing methods in the literature which make use of replicability to assess the quality of clustering analyses. 

\par One of the most important paradigms for the replicability of the analysis of complex, high-dimensional data follows under the notion of statistical stability \citep{yu2013stability,lim2016estimation,murdoch2019definitions,arrieta2020explainable}. Namely, the idea that a statistical analysis is replicable if it stable, i.e.\ it produces similar results when performed several times, using the same or slightly different data. Here, we discuss stability in the context of clustering analyses.

\subsubsection{Global Replicability} 

A simple measure of  clustering replicability  can be obtained by  comparing the partitions learned over multiple repetitions of algorithm $\AA$ on the same dataset $\Xtr$. If $\AA$ is stable, it should produce the same output  when re-run on the same input. Let $\psi^{(1)}, \psi^{(2)}$ be the two clustering functions learned by running $\AA$ twice on dataset $\Xtr$. It is possible that, even when repeatedly applied to the same dataset, multiple runs give raise to different results. E.g., if $\AA$ is $k$-means, distinct initializations --- potentially randomly selected --- may lead to different local minima of the objective function in \Cref{eq:opt_k_means}. This allows to think of $\psi$ as a random element.
Building on this idea, one can measure cluster replicability of $\AA$ by employing a measure of discrepancy or distance between the partitions of $\Xtr$ induced by $\psi^{(1)}$ and $\psi^{(2)}$. In this spirit, \citet{von2010clustering} suggested using  the minimal matching distance, defined as the minimum number of labels switches needed to make the partitions induced by $\psi^{(1)}$ and $\psi^{(2)}$ identical: 
\begin{equation}
	 \min_{\pi} \sum_{i=1}^n \ind\left[ \psi^{(1)}(x_i) \neq \pi \left\{\psi^{(2)}(x_i) \right\}\right], \label{eq:mmd}
\end{equation}
where $\pi$ is a permutation of the $k$ labels of the clusters. Notice that, while in this example we focused on $k$-means, the same argument could be applied to any other clustering algorithm (e.g., hierarchical clustering, or even regression-based clustering), by virtue of our remark on ``Induced Clustering Functions'' above.

A clustering algorithm could produce the same output when re-applied to the same dataset,  but its output might change considerably if we  just slightly change the input data. 
Several authors have therefore generalized the definition of stability by comparing the results of  an algorithm  applied to slightly different versions of the same dataset \citep{bryan2004problems, lange2004stability, ben2007stability}. This notion of clustering stability is widely accepted among practitioners, since a replicable clustering procedure should not be too sensitive to small perturbations of the data. Standard approaches to produce perturbed versions of the original dataset and to perform stability analyses include (i) sub sampling the data \citep{levine2001resampling}, or (ii) corrupting individual datapoints, e.g.\ by adding random noise  \citep{hennig2007cluster}. If the results obtained by performing clustering on the corrupted datasets are similar to the ones obtained on the original data,  then the cluster analysis is stable.   Instead, when the results differ   across  perturbed datasets --- despite the fact that the datasets are  similar by construction --- the  clustering  algorithm  is deemed unstable. 

A general stability-based measure of clustering replicability on dataset $\Xtr$ is then as follows. Let $B$ be a large integer. For $b=1,\ldots,B$ let $\Xtr^{(b)}$ be either  (i) a random subsample of $\Xtr$ (e.g., a draw of $n'<n$ datapoints without replacement), (ii) a corrupted version of $\Xtr$, in which the $i$-th datapoint $x^{(b)}_{i} = x_i + \epsilon_i^{(b)}$, where $\epsilon_i^{(b)}$ is a random error term or (iii) a combination of (i) and (ii).  A stability measure is obtained by  averaging
the minimal matching distance of \Cref{eq:mmd} between the partition  of $\Xtr$ learned using the full dataset and the partition obtained with the $b$-th corrupted dataset, 
\[
	 B^{-1}\sum_{b=1}^B d\left\{\Psi(\Xtr; \mathcal{A},\Xtr);  \Psi(\Xtr; \mathcal{A},\Xtr^{(b)}) \right\}.
\]

\subsubsection{Local Replicability} \label{sec:stability_local}

Besides assessing the global replicability of a clustering algorithm, scientists might be interested in understanding the {local} replicability of a specific cluster of interest, e.g.\ because this cluster is hypothesized to be biologically relevant. To do so, \citet{smolkin2003cluster} propose the following procedure:
run a clustering algorithm on $\Xtr$ and let $U_1,\ldots,U_k$ be the clusters identified.  Given a fraction $\alpha  \in (p^{-1},1]$ and a large integer value $B$, for $b=1,\ldots,B$, select a random subset of $p' = \lfloor \alpha p \rfloor$ covariates  from the $p$ original ones. Let $\Xtr^{(b)}$ be the dataset obtained by retaining for every datapoint only the $p'<p$ randomly selected covariates.  Perform the same clustering procedure as before, but now on  $\Xtr^{(b)}$, and let $U^{(b)}_1, \ldots, U^{(b)}_k$ be the resulting clusters. Stability of the $j$-th cluster $U_j$ is measured by the fraction of repetitions for which there exists a cluster $U^{(b)}_{\ell}$ such that $U_j \subseteq U^{(b)}_{\ell}$: 
\[
	\frac{1}{B}\sum_{b=1}^{B} \ind\left( \sum_{\ell=1}^k\ind(U_j \subseteq U_{\ell}^{(b)}) >0 \right).
\]
Variations of this methods and similar ideas have also been proposed. E.g.\ \citet{hennig2007cluster} uses the Jaccard coefficient: $\frac{1}{B}\sum_{b=1}^{B}\max_{i=1,\ldots,k}  \frac{|U_j \cap U_i^{(b)}|}{|U_j \cup U_i^{(b)}|}$. In either case, the mean across re-runs could be replaced by other summaries, such as the median or quantiles.

In another notable approach, \citet{mcshanerepro} propose two metrics for local clustering stability, the $R$-index and the $D$-index. The method relies on considering again $B$ perturbations $\Xtr^{(1)},\ldots,\Xtr^{(B)}$ of the original dataset $\Xtr$, with $\Xtr^{(b)} = \{x_1+e_1^{(b)},\ldots,x_n+e_n^{(b)}\}$, where $e_i^{(b)}$ are i.i.d.\ mean-zero Gaussian error terms with variance adequately chosen for the data under consideration. For  $U_j$ a cluster of interest in $\Xtr$, the $R$-index quantifies $U_j$'s stability by computing the average fraction of pairs of datapoints in $U_j$ which remains  clustered together after re-clustering  the perturbed dataset $\Xtr^{(b)} $ across re-runs. The $D$-index computes across re-runs the average number of  ``discrepancies'' (additions or deletions) between $U_j$  and the cluster with the highest overlap in $\Xtr^{(b)}$.

\subsection{Cluster Analyses and Prediction Accuracy} \label{sec:prediction_accuracy}

\subsubsection{Global Replicability}
A different approach for clustering replicability is driven by the idea of prediction accuracy, where concepts developed in the context of classification are adapted to clustering. In influential work \citet{tibshirani2005cluster} relied on the idea of prediction accuracy to develop a procedure for identifying the ``best'' clustering algorithm $\AA$ for the data at hand. In a nutshell, the ``best'' algorithm is the one that allows to  predict  with the highest accuracy clustering co-membership of points in a test set, using a clustering function learned on a training set. More precisely, first split the data into a training and test dataset $\Xtr = \{x_1,\ldots,x_n\}, \Xte = \{x_1',\ldots,x_m'\}$. Here, for simplicity, assume that $\AA$ identifies $k$ clusters. Then, 
\begin{itemize}
	\item Run $\AA$ on $\Xtr$ and on $\Xte$ separately. 
	 Let $U_1,\ldots,U_k$ denote the subsets of the partition of $\Xte$ induced by $\Psi(\Xte; \AA,\Xte)$, and $m_j = |U_j|$, the size of subset $j$. 
	\item Quantify the ``prediction strength''  by computing co-clustering occurrences: for any two points in the test set belonging to the same cluster, $x'_i, x'_{\ell} \in U_j$, let $\eta(x_i', x_j') =  \ind \left\{ \psi(x'_i; \AA,\Xtr) = \psi(x'_{\ell}; \AA,\Xtr)\right\} $. The prediction strength is defined as
	\begin{align}
		\textrm{ps}(k):= \min_{1\le j\le k} \left[  \frac{ \sum_{x'_i, x'_{\ell} \in U_j, i \neq \ell} \eta(x_i',x_j')}{{m_j(m_j-1)}} \right]. \label{eq:ps}
	\end{align}
\end{itemize} 
The authors use prediction strength to identify the number of clusters  $k$ into which the dataset should be partitioned --- namely, letting be  $k  = \arg\max_{k'\ge 2} \textrm{ps}(k') $.

\subsubsection{Local Replicability}
Relatedly, \citet{kapp2006clusters} developed a  cluster-specific measure of replicability, the ``in-group-proportion'' [IGP]. For a given cluster $U_j$, the IGP is the fraction of datapoints in $U_j$ whose nearest neighbors are also in the same group: let $\textrm{nn}(x_i) = \arg\min_{x' \in \Xtr} d(x_i, x')$, for  a  distance $d(\cdot,\cdot)$. Then,
\[
	\textrm{IGP}(U_j) = \frac{1}{|U_j|} \sum_{x_i \in U_j} \ind \left[ {\textrm{nn}(x_i)} \in U_j\right].
\]
This measure can be used to assess the replicability of an individual cluster $U_j$.

\subsection{Analyses Based on Tests of Statistical Significance} \label{sec:stat_tests}

Last we consider methods to assess cluster replicability via tests of statistical significance. At a high level, these work by building statistical tests for the ``null'' hypothesis $H_0$ that the data does not contain distinct clusters. 
Tests  can  provide  evidence  that  the data includes  two  or  more  clusters by rejecting the null hypothesis. Most tests for clustering replicability are parametric, i.e.\ they assume that under $H_0$ the data originates from a posited  parametric  model. The tests work by comparing the value of a relevant  test statistic,  computed using the observed data $\Xtr$, to the distribution of  the same statistic under the null hypothesis. If the value of the test statistic is sufficiently unlikely under the null, then $H_0$ is rejected. Following \citet{mcshanerepro}, we now discuss a general recipe to assess replicability of clustering using statistical tests:
\begin{enumerate}
	\item Let $F_0$ be a distribution from which data is drawn under $H_0$ (e.g., Gaussian). 
	\item Let $d_i= \min_{j\neq i} \| {x}_i - {x}_j \|^2$ be $x_i$'s nearest neighbor distance and 
	 $G_{\star}(v) = \sum_i \ind(d_i \le v)/n$  the cumulative density function (CDF) of $d_1,\ldots,d_n$. 
	\item Generate a sequence of datasets from the null model, ${\Xtr}^{(b)}$, for $b=1,\ldots,B$. 
	For each simulated dataset, obtain the CDF of the nearest neighbor distances $G_b(\cdot)$. 
	Under $H_0$, $G_{\star}, G_1,\ldots,G_B$  are  approximately  identically  distributed. 		
	\item Compute for $b \in \{\star,1,\ldots,B\}$ the test statistics:
	\[
		s_b = \bigintsss_0^{\infty} \left\{ G_{b}(y) - \frac{1}{B}\sum_{b' \neq b} G_{b'}(y) \right\}^2 \mathrm{d} y.
	\]
	Each $s_b$  can be  interpreted as  the distance  of $G_b(\cdot)$  from the average of the other CDFs, $\frac{1}{B}\sum_{b' \neq b} G_{b'}(\cdot)$. 
	\item Compare $s_\star$ to  $s_1,\ldots,s_B$. Reject  $H_0$ at confidence level $\alpha$ if  $s_\star$ is larger than the $100\times (1-\alpha)$  percentile of  $s_b$, $b=1,\ldots,B$. 
\end{enumerate}
Other measures besides the nearest neighbor distance (step 2) and other test statistics (step 4) could be employed (see  \citet{levenstien2003statistical,alexe2006data,bertoni2007model,liu2008statistical} for other approaches).

\subsection*{Available Software} \label{sec:review_code}

Among existing software for clustering replicability, we recommend the $R$ packages \texttt{clValid} \citep{brock2008clvalid} and \texttt{fpc} \citep{hennig2015package}, which support a few of the replicability indices discussed above. We provide code to generate and test all the replicability metrics discussed in \Cref{sec:review}. Experimental results for all the methods discussed are presented in \citet[Appendix B]{masoero2022clusteringsupp}. 

%% file: sec-theory_multi.tex
% !TEX root = main.tex

%%%%%%%%%%%%%%%%%%%%%%%%
\section{Clustering and Cross-Study Replicability} \label{sec:theory_multi}
%%%%%%%%%%%%%%%%%%%%%%%%

\subsection{Challenges and Strategies for Cross-Study Clustering Replicability}

Often, scientists interested in investigating replicability have access to \emph{multiple} datasets, and want to understand replicability properties of their analyses \emph{across} these datasets. Different datasets can have their own specificity,  for  example  because  of  different study designs or technical differences in the instrumentation used to measure the variables. Hence, high  replicability  scores within each study   do  not   necessarily imply   replicability  across different studies. Nonetheless it is possible to make progress in the more challenging assessment of replicability across multiple studies by extending some of the principles reviewed in the previous section.  Here we provide a guide for this extension, discussing how to leverage the ideas presented in \Cref{sec:review} for quantifying {cross-study} clustering replicability when {multiple studies} are available, and for selecting  clustering algorithms  with better  {cross-study}  replicability  properties  compared to  others. 

To get started, we need an operational definition for cross-study replicability. For simplicity,  consider two datasets $\Xtr, \Xte$ --- the ``training'' and ``testing'' dataset respectively. These could be data from independent studies probing the same molecular features on patients with similar clinical conditions. Informally, we say that clustering algorithm $\AA$ is replicable across $\Xtr$ and $\Xte$ if it is able to learn similar clustering functions across these two studies.  We quantify the similarity of the clustering functions learned on $\Xtr$ and $\Xte$ in an intuitive way, by measuring the differences  between the partitions of $\Xte$ obtained by using the clustering functions learned by training $\AA$ on  $\Xtr$   and on $\Xte$ respectively. That is, by computing $d(\Psi(\Xte;\AA,\Xtr), \Psi(\Xte; \AA, \Xte))$, where $d(\cdot,\cdot)$ is a metric of discrepancy between two partitions of the  set $\Xte$. 

Notice that this setting is similar to what already discussed in \Cref{sec:prediction_accuracy}: e.g., the prediction strength of \Cref{eq:ps}, is an example of a similarity metric on partitions. However, while in the discussion in \Cref{sec:prediction_accuracy} the train and test data are random subsamples from the same dataset, here we allow for $\Xtr$ and $\Xte$ to be different, independent datasets. As a consequence of that, differently from the case discussed in \Cref{sec:prediction_accuracy}, we do not expect a priori $d(\cdot,\cdot)$ to be symmetric in its arguments.

 In our discussion, we use cancer subtype validation as a motivating example: we consider multiple datasets of patients gene expressions, collected by different investigators.  Replicability analysis will help us understand whether the clustering learned on the dataset collected by one investigator identifies cancer subtypes in a different dataset.

\subsection{Useful Metrics for Clustering Replicability}

A key ingredient to quantify cross-study replicability is the choice of discrepancy metric $d(\cdot, \cdot)$ between partitions. Several options exist in the literature \citep{albatineh2006similarity, vinh2010information, jaskowiak2014selection}. Here, we consider two metrics: the Rand index (RI, \citet{rand1971objective}) and the mutual information (MI). In what follows, let $\bm{U}=\{U_1,\dots,U_{k_1}\}$ and $\bm{V} = \{V_1,\dots,V_{k_2}\}$ denote two partitions of the set $\Xte=\{x'_1,\ldots,x'_m\}$, with $|\bm{U}| = k_1$ and $|\bm{V}| = k_2$, respectively. 

 \paragraph*{The Rand Index [RI]} The Rand index is a simple pair-counting discrepancy measure, obtained  by computing the fraction of pairs of datapoints on which two  partitions $\bm{U}$ and $\bm{V}$ of the same set $\Xte$ agree. Let $N_{0,0}$ denote the number of pairs of points which belong to different subsets in both partitions, $N_{1,1}$  the number of pairs which belong to the same subset in both partitions, $N_{1,0}$  the number of pairs which belong to the same subset in partition $\bm{U}$, but are in different subsets in partition $\bm{V}$, and symmetrically $N_{0,1}$ the number of pairs belonging to different subsets in partition $\bm{U}$ and to the same subset in partition $\bm{V}$. Because there are exactly $\binom{m}{2}$ pairs of points, and each pair of points falls in exactly one of these categories, it follows that
\[
	N_{0,0}+N_{0,1}+N_{1,0}+N_{1,1} = \binom{m}{2}.
\]
 The Rand index for partitions $\bm{U}$, $\bm{V}$ of the set $\Xte$ is then defined as
\begin{align} \label{eq:RI}
	\RI(\bm{U},\bm{V}) = \frac{N_{0,0}+N_{1,1}}{\sum_{i=0}^1\sum_{j=0}^1 N_{i,j}}= \frac{N_{0,0}+N_{1,1}}{\binom{m}{2}}. 
\end{align}

\paragraph*{The Mutual Information [MI]} The mutual information is an information theoretic quantity which provides a measure of the dependence between two random variables. In clustering, the mutual information between two partitions quantifies how much information about one  cluster  membership is revealed by knowing the other partition. The mutual information between partitions $\bm{U}, \bm{V}$ of $\Xte$ is
\begin{align} \label{eq:MI}
	\MI(\bm{U}, \bm{V}) &= \sum_{i=1}^{k_1}\sum_{j=1}^{k_2} \frac{| U_i \cap V_j|}{m} \log \left(\frac{\frac{| U_i \cap V_j|}{m}}{\frac{|U_i|\times|V_j|}{m^2}}\right).
\end{align}
Given clustering functions $\psi(\cdot; \Xtr)$ and $\psi(\cdot; \Xte)$, we use the Rand index $\RI(\Psi(\Xte; \Xtr), \Psi(\Xte; \Xte))$, or the mutual information $\MI(\Psi(\Xte; \Xtr), \Psi(\Xte; \Xte))$ as  replicability metric of the clustering learned on $\Xtr$ and tested on $\Xte$. 

\paragraph*{Adjustments to Improve Interpretability} A desirable property that should be shared across  replicability metrics is that they should yield, at least in expectation, a constant and common baseline value when applied to independent random partitions. That is to say, if we  repeatedly create independent random partitions $\bm{U}$ and $\bm{V}$ of  $\Xte$, then on average the replicability measure chosen should be a baseline, constant value.
This is generally not the case for either the Rand index or the mutual information. To overcome this shortcoming, we adopt the standard solution of \citet{hubert1985comparing}, and derive ``adjusted'' indices  satisfying the property above when applied to independent random clusterings. First, we fix a probabilistic model according to which random partitions are generated. We follow the standard convention and choose the permutation model \citep{lancaster1969}, in which clusterings are generated at random under the constraint of a fixed number of clusters, and a fixed number of points within each cluster. Under this model, we define the following adjusted metric:
\[
	r^\star(\bm{U}, \bm{V}) = \frac{r(\bm{U}, \bm{V}) - \EE_{\tilde{\bm{U}}, \tilde{\bm{V}}}[r(\tilde{\bm{U}}, \tilde{\bm{V}})]}{\max_{\bm{U}',\bm{V}'}\{r(\bm{U}', \bm{V}')\} - \EE_{\tilde{\bm{U}}, \tilde{\bm{V}}}[r(\tilde{\bm{U}}, \tilde{\bm{V}})]},
\]
where $r(\bm{U}, \bm{V})$ denotes the unadjusted replicability metric (RI or MI) between partitions $\bm{U}$ and $\bm{V}$ of the same dataset $\Xte$. Given $\bm{U}$ and $\bm{V}$ and their subsets' sizes, the terms $\EE_{\tilde{\bm{U}}, \tilde{\bm{V}}}[r(\tilde{\bm{U}}, \tilde{\bm{V}})]$ and $\max_{\bm{U}',\bm{V}'}\{\bm{U}', \bm{V}'\}$ appearing in the denominator of the equation above are obtained by respectively integrating and maximizing over the set of partitions  with $k_1$ and $k_2$ subsets having sizes $\{|U_k|\}_{k=1}^{k_1}$ and $\{|V_j|\}_{j=1}^{k_2}$ with respect to the permutation model of \citet{lancaster1969}. We henceforth adopt the adjusted replicability indices described above: the adjusted Rand index (ARI) and mutual information (AMI), and replace them in \Cref{eq:RI,eq:MI}. For additional details on this approach, see \cite{vinh2009information, vinh2010information}.  

%%%%%%%%%%%%%%%%%%%%%%%%%%
\subsection{Quantifying Cross-Study Replicability of Cluster Analyses} \label{sec:boot}
%%%%%%%%%%%%%%%%%%%%%%%%%%

%%%%%%%%%%%%%%%%%%%%%%%%%%
\subsubsection{Replicability on a Single Testing Dataset} \label{sec:repro_couple}
%%%%%%%%%%%%%%%%%%%%%%%%%%

Let $\Xtr, \Xte$ be independent datasets consisting of $n$ and $m$ data points respectively. To capture  heterogeneity caused by different measurement technologies, errors in the measurement processes, and other factors, we model each dataset as a collection of independent and identically distributed draws from separate distributions $F_1$ and $F_2$ with support on $\R^p$. Our goal is to evaluate the replicability of clustering algorithm $\AA$ trained on data $\Xtr$ drawn from $F_1$ and  validated on data $\Xte$  from $F_2$. To capture it, we  define the cross-study cluster analysis replicability index
\begin{align}\label{eq:repro_1}
	R(F_2; \AA, F_1 ) = \EE[r^\star(\Psi(\bm{Z}'; \AA, \bm{Z}), \Psi(\bm{Z}'; \AA,\bm{Z}'))],
\end{align}
where $\bm{Z} = \{Z_1,\ldots,Z_n\}$ is a collection of $n$ i.i.d.\ random replicates from $F_1$, and $\bm{Z}'=\{Z_1',\ldots,Z_m'\}$ is a collection of $m$ i.i.d.\ random replicates from $F_2$. We refer to the index in \Cref{eq:repro_1} as $R$. 

Notice that this index depends implicitly also on the sample sizes $n, m$ of the training and testing datasets. Indeed, useful replicability metrics will typically depend on sample size, as we discuss in \citet[Appendix A.3]{masoero2022clusteringsupp}. A point estimate of $R$ is obtained via $\hat{R}:=r^\star(\Psi(\Xte; \AA, \Xtr), \Psi(\Xte; \AA, \Xte))$. 

To produce interval estimates of $R$, we employ a bootstrap approach, similar to those proposed for within-study performance (see, e.g.\ \citet{fang2012selection}). We fix a number $B_1$ of bootstrap replicates and for each $b\in[B_1]$, we generate bootstrap datasets $\Xtr^{(b)}, \Xte^{(b)}$ by sampling with replacement $n$ and $m$ data points from $\Xtr$ and $\Xte$ respectively. For every $b\in[B_1]$, we estimate the replicability score $\hat{R}^{(b)}:=r^\star(\Psi(\Xte^{(b)}; \AA, \Xtr^{(b)}), \Psi(\Xte^{(b)}; \AA, \Xte^{(b)}))$. The values $\{\hat{R}^{(1)},\ldots,\hat{R}^{(B_1)}\}$ yield an estimate the variability of    $\hat{R}$. We discuss calibration of these estimates in \citet[Appendix A.1]{masoero2022clusteringsupp}, and summarize our procedure in \Cref{algo_1}.

%%%%%%%%%%%%%%%%%%%%%%%%%%
\subsubsection{Replicability Across a Collection of Datasets}  \label{sec:rep-index}
%%%%%%%%%%%%%%%%%%%%%%%%%%
We next  consider the scenario in which a collection  $\XX = \{\Xtr_1,\dots,\Xtr_S\}$ of $S>2$ datasets is available. Each $\Xtr_s$ contains $n_s$ samples of the same $p$ features. Extending the ideas of \Cref{sec:repro_couple}, we define the cross-study replicability of algorithm $\AA$ trained on dataset $\Xtr_1$  and tested on $\Xtr_{s'}$, $s' = 2,\ldots,S$ as the average over pairwise replicability scores:
\begin{align}
	\mR(\{F_{s'}\}_{s'=2,\ldots,S}; \AA, F_1) = \sum_{s' \neq 1} \frac{R(F_{s'}; \AA, F_1)}{S-1},
\end{align}
with $R(F_{s'}; \AA, F_s)$ defined in \Cref{eq:repro_1}. Simliar expressions can be defined for training sets other than $\Xtr_1$. A point estimate $\hat{\mR}$ of $\mR(F_{s'}\}_{s'=2,\ldots,S}; \AA, F_1) $ is obtained by replacing each score $R(F_{s'}; \AA, F_1)$ with its sample counterpart $\hat{R}_{s'}:=r^\star(\Psi(\Xtr_{s'}; \AA, \Xtr_1), \Psi(\Xtr_{s'}; \AA, \Xtr_{s'}))$. We adopt the same approach as before to produce interval estimates of ${\mR} $. Fix a number $B_2$ of iterations, and for each $b \in [B_2]$ draw a bootstrap copy $\Xtr_{s}^{(b)}$ from $\Xtr_s$, $s \in [S]$. We learn clustering functions $\psi(\cdot; \AA, \Xtr_s^{(b)})$, $s \in [S]$, and compute for each $s'=2,\ldots,S$ the  replicability index $\hat{R}_{s'}^{(b)}:= r^\star(\Psi(\Xtr_{s'}^{( b)}; \AA, \Xtr_{1}^{(b)}), \Psi(\Xtr_{s'}^{(b)}; \AA, \Xtr_{s'}^{(b)}))$. 
Then, for every $b  \in[B_2]$, compute $\hat{\mR}^{(b)} :=(\hat{R}_{2}^{(b)} + \ldots + \hat{R}_{S}^{(b)})/(S-1)$. The values $\{\hat{\mR}^{(1)}_1,\dots, \hat{\mR}^{(B_2)}_1 \}$ are used to produce intervals around $\hat{\mR}(\{\Xtr_{s'}\}_{s'=2,\ldots,S}; \AA, \Xtr_1)$, as described in \Cref{algo_2}.
\begin{algorithm}
\caption{}  \label{algo_1} 
    \begin{algorithmic}[1]  
	 \Procedure{}{} Fix clustering algorithm $\AA$, replicability index $r$, number of bootstrap iterations $B_1$. Let $\Xtr$ and $\Xte$ be training and testing datasets. 
            \For{$b \in 1,\ldots,B_1$}
                \State Draw bootstrap samples $\Xtr^{(b)} , \Xte^{(b)} $ from $\Xtr, \Xte$.
                \State Learn $\psi(\cdot; \Xtr^{(b)})$ and $\psi(\cdot; \Xte^{(b)})$.
                \State Compute $\hat{R}^{(b)} = r^\star(\Psi(\Xte^{(b)}; \Xtr^{(b)}), \Psi(\Xte^{(b)}; \Xte^{(b)}))$.
            \EndFor            \State \textbf{return} $\hat{R}^{(1)},\dots, \hat{R}^{(B_1)}$.
        \EndProcedure
    \end{algorithmic}
\end{algorithm}
\begin{algorithm} 
\caption{}  \label{algo_2} 
    \begin{algorithmic}[1] % The number tells where the line numbering should start
        \Procedure{}{}            Fix clustering algorithm $\AA$, replicability index $r$, number of bootstrap iterations $B_2$, training dataset $\Xtr_t$, $t\in[S]$. Let $\XX = \{X_1,\ldots,X_S\}$ be the collection of available datasets.
            \For{$b = 1,\dots, B_2$}
            	\For{$s=1,\dots,S$}
            		\State  Draw bootstrap sample $\Xtr_{s}^{(b)}$ from $\Xtr_s$. 
			\State Learn $\psi(\cdot; \Xtr^{(b)}_s)$.
		\EndFor
		\For{$s = 1,\ldots, S ,\quad  s \neq t$}
			\State Let $\hat{R}_{s}^{(b)} = r^\star(\Psi(\Xtr_{s}^{(b)}; \Xtr_{t}^{(b)}), \Psi(\Xtr_{s}^{(b)}; \Xtr_{s}^{(b)}))$.
		\EndFor
		\State Let $\hat{\mR}^{(b)} = \frac{1}{S-1} \sum_{s\neq t}\hat{R}^{(b)}_{s}$.
            \EndFor
            \State \textbf{return} $\hat{\mR}^{(1)},\dots, \hat{\mR}^{(B_2)}$.
        \EndProcedure 
    \end{algorithmic}
\end{algorithm}

\subsubsection{Local Replicability: Validating Individual Clusters} \label{sec:ind_cl}

The replicability indices produced by \Cref{algo_1,algo_2} provide measures to quantify which clustering procedures  are reproducible  across multiple datasets at a \emph{global} scale. A natural additional desideratum for a reproducible clustering procedure is that it will induce partitions that are similar also locally. Indeed, sometimes, the clusterings found on two or more datasets can agree on the majority of the sample space, leading to high replicability scores, but disagree on a smaller portion.  For example, when two datasets are available, one clustering might completely miss a small cluster that is clearly identified by the other. 
The replicability scores  provided  by \Cref{algo_1,algo_2} can fail to capture this pattern. Conversely, it may be that the two clusterings disagree almost everywhere, but are able to identify one or a few robust clusters.  In this case, the replicability scores might be poor, and fail to provide insight about which clusters are reproduced across datasets. In applications, measuring which clusters replicate across datasets is important, as it can provide insights about the presence or absence of specific groups of practical interest.

As before, let $\Xtr$, $\Xte$ be the training and testing datasets, from distributions $F_1$ and $F_2$ respectively. We now present a strategy to leverage the replicability scores when one is interested in understanding local replicability, with respect to a point of interest $x\in \R^p$. For example, $x$ can be a specific patient clinical profile.  We can quantify whether clustering algorithm $\AA$ produces similar results in the neighborhood of $x$ when trained on data $\Xtr$ and tested on data $\Xte$ via the following ``local replicability score'' ($\rm{LR}$):
\begin{equation*} 
	\textrm{LR}(F_2; \AA, F_1, x)=  \EE[r^\star(\tilde{\Psi}_x(\bm{Z}'; \AA, \bm{Z}), \tilde{\Psi}_x(\bm{Z}'; \AA, \bm{Z}'))].
\end{equation*}
Notice that the local replicability score $\textrm{LR}$ is simply obtained by replacing the clustering function $\psi$ used in \Cref{eq:repro_1} with the binary clustering operator $\tilde{\psi}_x$. If $x$ belongs to a reproducible cluster, then the two binary partitions should be similar, and the local replicability score high. Conversely, if $x$ is a data point within a cluster with low reproducibility, these binary partitions differ, and local replicability scores will be low. We adopt the same approach used in \Cref{algo_1,algo_2} to obtain bootstrap confidence intervals (see \Cref{algo_3} for details).

\paragraph*{Cluster-Specific Replicability} The procedure outlined above quantifies replicability with respect to a single point, but the same strategy can be used to assess replicability at a cluster level. For example, to assess the replicability of cluster $U_1 \subset \Xtr$, we average the local replicability scores with respect to all points $x \in U_1$.

\paragraph*{Multiple Studies} When multiple studies $\Xtr_1,\ldots,\Xtr_S$ are available and we are interested in studying the local replicability of a clustering learned on a training set $\Xtr_s$ when validated against $\Xtr_t$, $t\neq s$, we can simply perform \Cref{algo_3} for all pairs of datasets $(s,t)$, $t=1,\ldots,s-1,s+1,\ldots,S$.
\begin{algorithm}
    \caption{} \label{algo_3}
    \begin{algorithmic}[1]  
        \Procedure{}{} Fix clustering algorithm, replicability index $r$, number of bootstrap iterations $B_3$, $x \in \R^p$.
            \For{$b = 1,\dots, B_3$}
               	\State Draw bootstrap samples $\Xtr^{(b)}$ from $\Xtr$, $\Xte^{(b)}$ from $\Xte$. 
	 	\State Learn $\psi(\cdot; \Xtr^{(b)})$ and $\psi(\cdot; \Xte^{(b)})$.               	
		\State{Let $\widehat{\textrm{LR}}^{(b)}=r^\star(\tilde{\Psi}_x(\Xte^{(b)}; \Xtr^{(b)}), \tilde{\Psi}_x(\Xte^{(b)}; \Xte^{(b)}))$}
            \EndFor
            \State \textbf{return} $\widehat{\textrm{LR}}^{(1)},\ldots,\widehat{\rm{LR}}^{(B_3)}$
        \EndProcedure
    \end{algorithmic}
\end{algorithm}

%% file: sec-exp-synth.tex
% !TEX root = main.tex
\section{Simulation Study} \label{sec:exp-synth}
\subsection{Homogeneous Datasets } \label{sec:exp_algo_1}

To demonstrate the usefulness of the replicability metrics introduced, we start by showing, on synthetic data, how the replicability index can guide the choice of the number of clusters and clustering algorithm. We consider four popular clustering algorithms: Birch \citep{zhang1996birch}, $k$-means, mini-batch $k$-means \citep{lloyd1982least}, and agglomerative clustering \citep{ward1963hierarchical}. All these procedures take the number of  clusters $k$ as an input. Because we expect to observe high replicability scores when an appropriate algorithm and number of  clusters are chosen, we show how the replicability index can be used  to (i) tune the number of clusters for each algorithm and (ii) choose which algorithm to use.  

\begin{figure*}
\includegraphics[width = \textwidth]{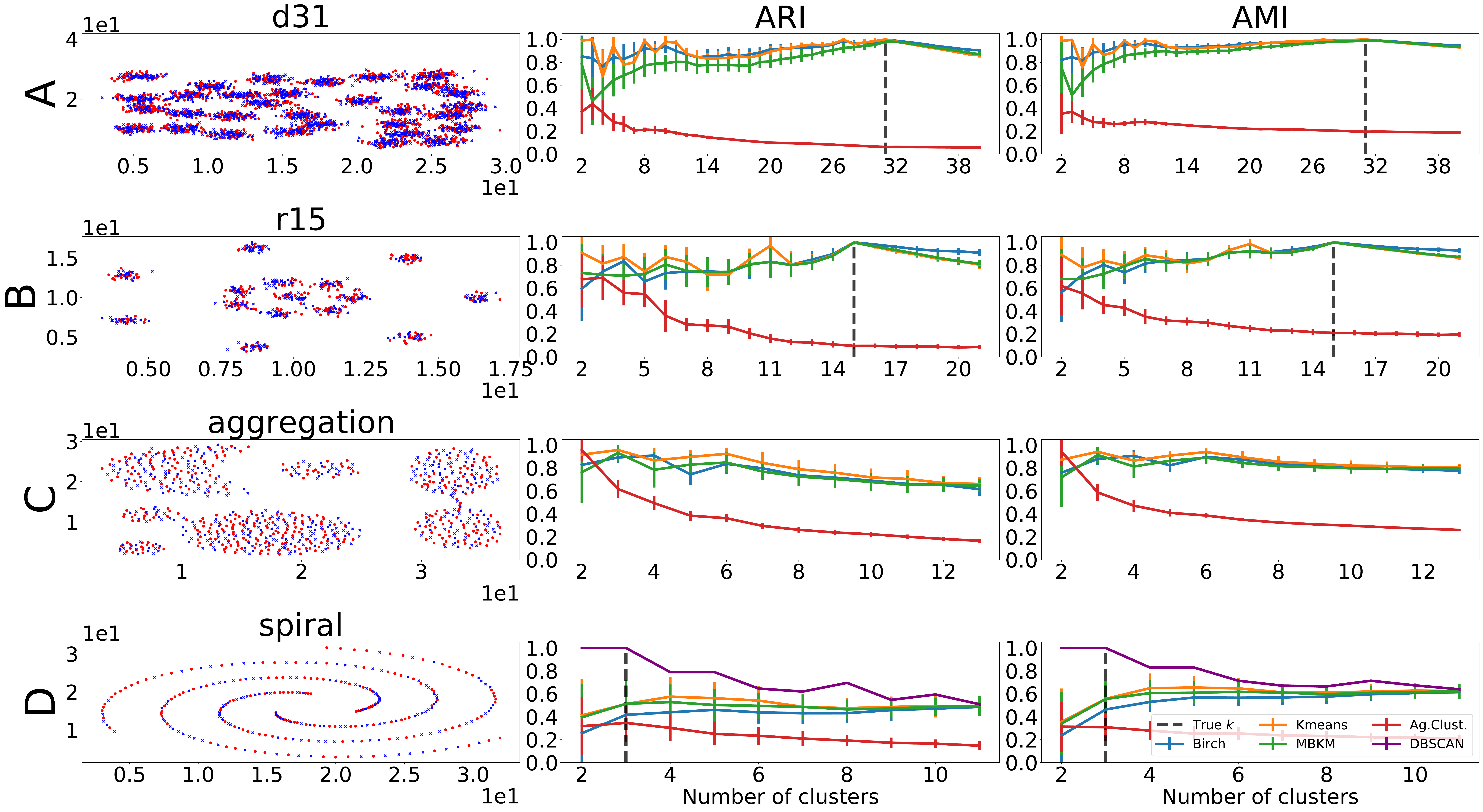}
    \caption{Reproducibility scores produced by \Cref{algo_1} on benchmark datasets \citep{ClusteringDatasets}. Left column: datasets (\emph{d31 (A), r15 (B), aggregation (C), spiral (D)}) randomly divided in training (red points) and testing (blue points) datasets of equal size. Center and right columns: replicability scores (vertical axis) using AMI (center) and ARI (right), versus number of clusters $k$ (horizontal axis). We used \Cref{algo_1} with $B = 100$ bootstrap iterations across different input values $k$ and clustering algorithms. We plot the mean score $\pm$ one standard deviation across bootstrap runs. The number of data points in datasets $A-D$ varies widely (d31: $n=3100$, r15: $n=600$, aggregation: $n=788$, spiral: $n=312$).}
    \label{fig:algo_1}
\end{figure*}

In our experiments,  we use  a collection of benchmark datasets (see the first column of \Cref{fig:algo_1}) \citep{ClusteringDatasets}. We partition each dataset into training and testing sets of equal size, and for each $k \in \texttt{range}_{40} = \{2,3,\dots,40\}$, we run \Cref{algo_1} for each clustering algorithm for $B_1=100$ bootstrap iterations. Our results are  illustrated   in \Cref{fig:algo_1}.  We highlight three important  features  of  the cross-study  replicability estimates obtained:
 \begin{itemize}
 	\item When the clusters are well separated, and the clustering algorithms used are appropriate for the shape of the clusters, the replicability indices achieve the  highest scores at the ``true'' number of  clusters  (see datasets $A$ and $B$).
	\item When the number of clusters is less obvious, as in dataset $C$, the replicability scores tend to be less peaked around a single value, indicating  uncertainty  on the  number  of  clusters  that  should be  selected to maximize replicability.
 	\item Sometimes, a clustering structure is present in the data, 
	but algorithms fail to capture it. For example, for dataset $D$,
	the only successful algorithm is DBSCAN (\citet{dbscan}, purple line), which achieves  good replicability scores for $k = 2,3$. In this case (D), the replicability index can aid the choice of the most appropriate clustering algorithm  for the data at hand.
 \end{itemize}

\subsection{Heterogeneous Datasets} \label{sec:exp_algo_2}
 To test \Cref{algo_2}, we consider datasets drawn from high-dimensional Gaussian mixture models. Through heterogeneity of the data generating distributions, we simulate differences we expect to observe across different groups of studies. A datapoint in the $s$-th study is drawn as follows:
\[
x_{s,i} \overset{i.i.d.}{\sim} F_{z_s} = \sum_{\ell=1}^k \pi_\ell^{(z_s)} \mathcal{N}\left(\mu_\ell^{(z_s)}, \Sigma_\ell^{(z_s)}\right),
\]
for $s = 1,\ldots,S $ and $i =1,\ldots, n_s$. Here, $z_s$ is an  index identifying the reference distribution from which dataset $s$ is generated, $\pi_\ell^{(z_s)} \in (0,1)$ are mixing proportions such that $\sum_\ell \pi_\ell^{(z_s)} = 1$ , and $\mathcal{N}(\mu_\ell^{(z_s)},\Sigma_\ell^{(z_s)})$ denotes  a Gaussian distribution with mean $\mu_\ell^{(z_s)}$ and covariance matrix $\Sigma_\ell^{(z_s)}$.
We set $S=4$ and assume $\bm{X}_1, \bm{X}_2$ are  drawn from the same mixture of Gaussians  $F_1$ (i.e. $z_1 =  z_2 =  1$), while $\bm{X}_3, \bm{X}_4$ are drawn from a second, different mixture of Gaussians $F_2$ (i.e.\ $z_3 = z_4 = 2$). Specifically, $F_1$ and $F_2$ are mixtures of Gaussians with  $k= 16$ components, each component with the same covariance matrix $\Sigma = I$, where $I$ is the  identity matrix. The mean vectors of $F_1$ are denoted by $\mu_{\ell}^{(1)} \in \R^{64}, \ell =1,\ldots, 16 $, and induce well separated components. The mean vectors of $F_2$ are denoted by $\mu_{\ell}^{(2)} \in \R^{64}, \ell =1,\ldots, 16$ and are obtained as follows: first a permutation $\pi(\cdot):[32]\to[32]$ of the first $32$ coordinates is chosen, and then each vector $\mu_{\ell}^{(2)}$ is obtained by permuting the first $32$ coordinates of  $\mu_{\ell}^{(1)}$ according to $\pi(\cdot)$. In this way, both $F_1$ and $F_2$ have $k=16$ well separated - albeit, different - components (see \citep{DIMsets} for further description of these distributions).

\begin{figure*}
        \centering \includegraphics[width=.8\textwidth]{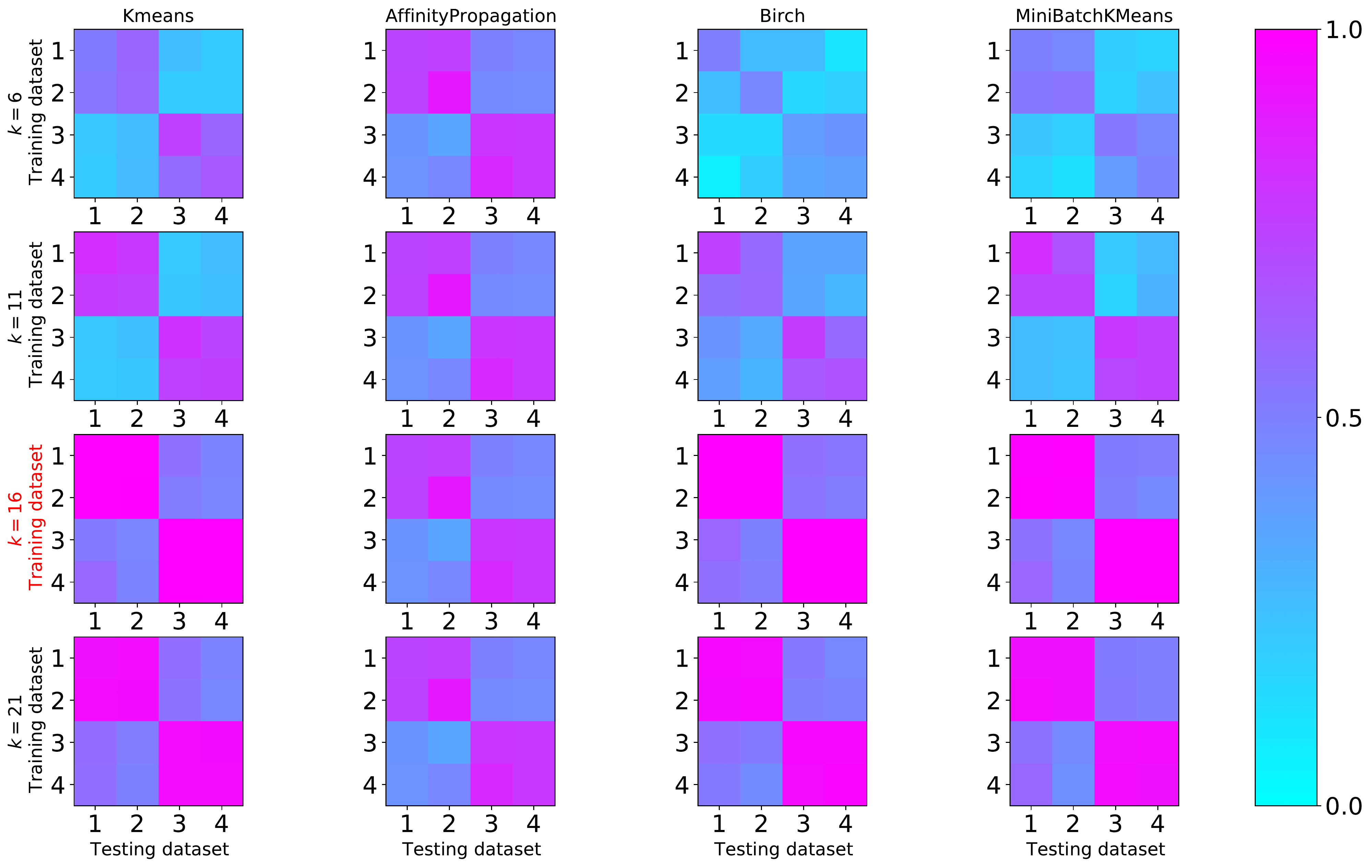}
        \caption{ARI scores for $4$ datasets in $\R^{64}$, $B=100$ bootstrap runs. Along the rows we vary the number of clusters $k \in \{ 6,11,16, 21 \}$, $16$ is the true number of clusters. Along the columns we vary the clustering algorithms.}\label{fig:algo_2}
\end{figure*}

 We compute and report, for every pair $\Xtr_s$ and $\Xtr_t$ of datasets, $s,t=1,\ldots,4$ the average replicability score  (\Cref{fig:algo_2}) for different clustering methods.  As expected,  replicability scores are always higher when comparing two datasets drawn from the same distribution, hence  sharing the same clustering structure (top left and bottom right block diagonals of each of the 16 sub-panels in \Cref{fig:algo_2}), while they are substantially lower when comparing datasets from different distributions. We also notice that all the clustering algorithms have better replicability when the correct number of clusters is specified as  input, and that  {\it Affinity propagation } \citep{frey2007clustering} seems to be performing overall worse than the other algorithms.
 
\subsection{Local Replicability}

\begin{figure*}
	 \includegraphics[width=\textwidth]{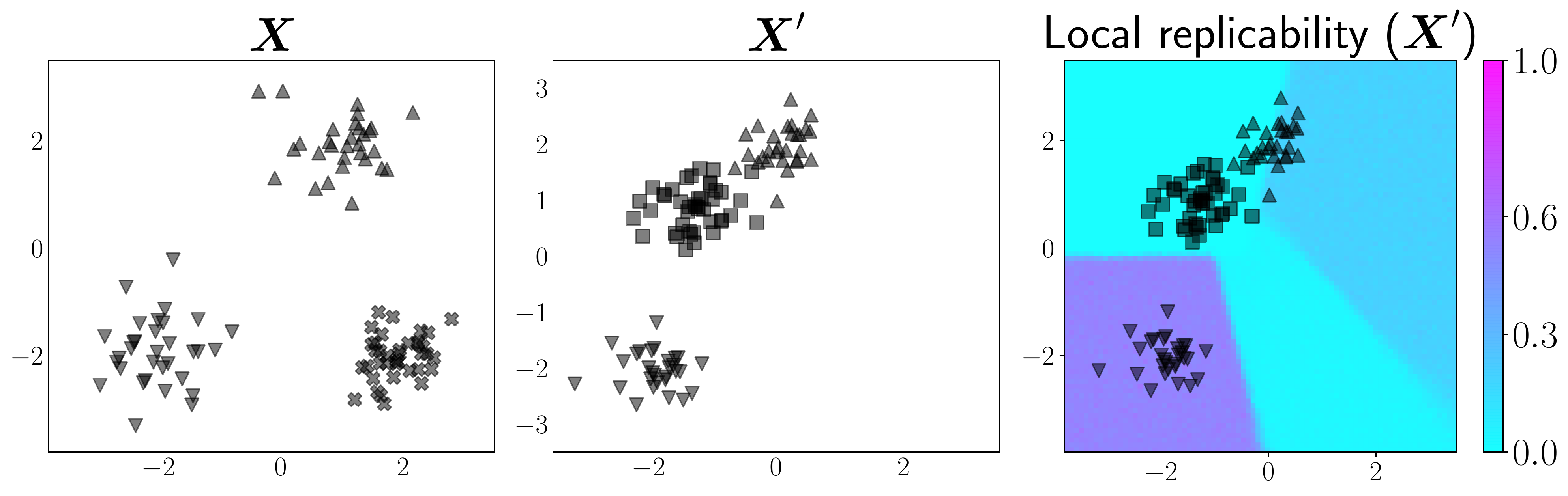} 
	  \caption{Local replicability. Left: training dataset $\bm{X}$ (from $F_1$). Center: testing dataset $\bm{X}'$ (from $F_2$, \Cref{eq:dist_local}). Different markers identify different components $(\triangledown: A, \triangle: B, \times: C, \square: D$). Right: local replicability on $\Xte$. We plot points in $\Xte$, and color the  sample space according to the local replicability scores of the partition learned on $\Xtr$ when tested on $\Xte$. Scores (ARI) are obtained by averaging $B_3 = 100$ bootstrap runs over a grid of $250 \times 250$ equally spaced points.} 
\label{fig:ind_cl}
\end{figure*}

To illustrate \Cref{algo_3}, we consider $S=2$ datasets, $\bm{X},\bm{X}'$ with $100$ datapoints each, drawn from Gaussian mixture models $F_1$ and $F_2$ in \cref{eq:dist_local} with three components and support on $\R^2$. $F_1$ and $F_2$ share two of the three components: letting $\mathcal{N}(\mu,\sigma^2)$ denote the distribution of a Gaussian random variable with mean $\mu$ and variance $\sigma^2$,

\begin{align} \label{eq:dist_local}
\begin{split}
	 F_1 =  \sum_{j \in \{A,B,C\}} \frac{1}{3}\mathcal{N}(\mu_j, \sigma^2 I), \\
	F_2=  \sum_{\ell \in \{A,B,D \}} \frac{1}{3} \mathcal{N}(\mu_\ell, \sigma^2 I).
\end{split}
\end{align}
We let $\sigma = 0.2$, $\mu_A = [ - 2, - 2 ]^\top$, $\mu_B = [ 0, 2 ]^\top$, $\mu_C = [2, -2]^\top$ and $\mu_D  = [-13/10, 13/20]^\top$. We refer to the mixture component identified by $\mu_A$ as component $A$, and similarly for the other components $B,C,D$. $\bm{X}$ has three well separated clusters, while two of the three clusters in $\bm{X}'$ are closer to each other (see \Cref{fig:ind_cl}).
 \begin{figure*}
 	\includegraphics[width=\textwidth]{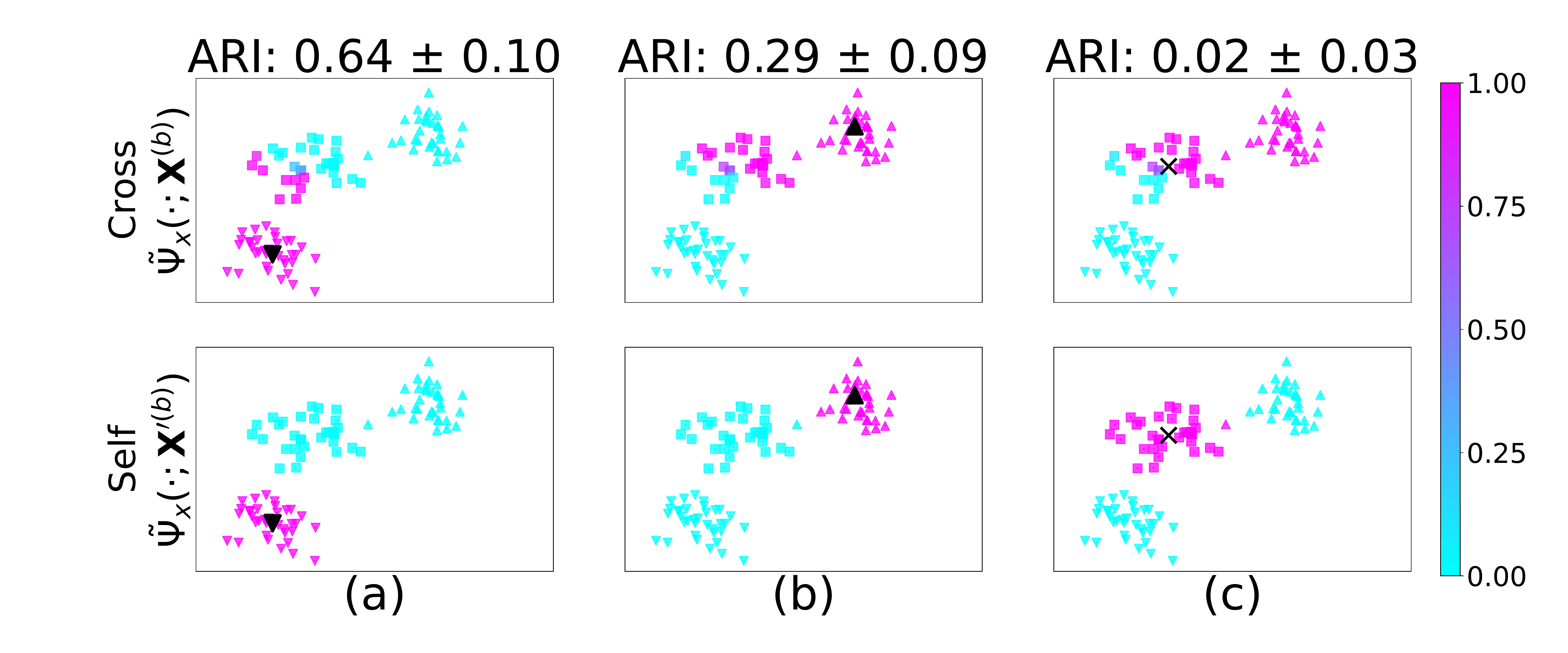} 
	\caption{Local replicability with $k$-means, $k=3$. Both rows show the test dataset $\Xte$. Column (a): we fix test point $x_A$ from component $A$ (in black). We color each data point $x' \in \Xte$ according to the average of $\{\tilde{\psi}_{x_A}(x'; \Xtr^{(b)})\}_{b=1}^{1000}$ (top row), $\{\tilde{\psi}_{x_A}(x'; \Xte^{(b)})\}_{b=1}^{1000}$ (bottom row). $\Xtr^{(b)}, \Xte^{(b)}$ are bootstrap draws from $\Xtr, \Xte$. We compute the LR (ARI) $\{r^\star(\tilde{\Psi}_{x_A}(\Xte^{(b)}; \Xtr^{(b)}), \tilde{\Psi}_{x_A}(\Xte^{(b)}; \Xte^{(b)}))\}_{b=1}^{1000}$, and report  the average score $\pm$ one standard deviation on top of column (a). This is high whenever the color pattern in the two rows is similar. Columns (b) and (c) replicate the analysis using points $x_B$ and $x_C$ from components $B$ and $C$.}   \label{fig:ind_cl_2}
 \end{figure*}

The goal here is to  quantify, at a fixed data point $x$, local replicability, as discussed in \Cref{sec:ind_cl}. For example, $x$ could be a point of $\bm{X}, \bm{X}'$, or any point in the support of the distribution. In our experiments,  \Cref{algo_3} is able to capture the different replicability properties of individual clusters, providing substantially different  scores for points belonging to different clusters. 
 
 We illustrate the  mechanics of the algorithm, and its ability to capture  local replicability  in \Cref{fig:ind_cl_2}: in column  (a), we consider a point $x_A$ close to  $\mu_A$, which does not belong to the    $\bm{X}$ and   $\bm{X}'$ datasets,
 and is identified by the $\triangledown$ symbol.  Component $A$ is present both in $\bm{X}$ and $\bm{X}'$, and the associated cluster replicates across the two datasets.
   Indeed, across the $B_3=100$ iterations, the two binary partitions $\tilde{\Psi}_{x_A}(\Xte^{(b)}, \Xtr^{(b)})$ and $\tilde{\Psi}_{x_A}(\Xte^{(b)}, \Xte^{(b)})$ (top, bottom row) are similar, and the replicability score relatively high.
In column (b), we consider instead a point $x_B$ belonging to the cluster induced by component $B$, identified by the $\triangle$ symbol.  In $\bm{X}$, this cluster is well separated from the clusters induced by the other components. However, in $\bm{X}'$, the cluster induced by component $B$ is close  to  the cluster induced by the component $D$, whose data points are identified  with the $\square$ symbol, affecting the replicability score associated with $x_B$. The  binary partitions  $\tilde{\Psi}_{x_B}(\Xte; \Xtr)$ and $\tilde{\Psi}_{x_B}(\Xte; \Xte)$ are different. $\tilde{\Psi}_{x_B}(\Xte; \Xtr)$ incorrectly indicates points from the  components $B$ and $D$ as belonging to the same cluster, while  the partition  $\tilde{\Psi}_{x_B}(\Xte; \Xte)$ does not incur into this problem. This discrepancy negatively affects the local replicability score. Last, in column (c), we consider a point $x_C$ that belongs to the cluster component $C$ in $\bm{X}$ ($\times$ symbol). In this case, the scores are essentially equal to $0$, since this cluster is absent in $\bm{X}'$. For the same data, we report results for the local replicability test scores at the cluster levels in \Cref{fig:ind_cl}.
  
%% file: sec-exp-real.tex
% !TEX root = main.tex
\section{Breast Cancer Gene Expression Datasets} \label{sec:exp-real}

\par In this section, we apply our clustering replicability metrics to publicly available breast cancer gene expression datasets that were collected to identify tumor subtypes. We consider the Mainz, Transbig and Vdx datasets \citep{mainz,transbig,vdx}. These datasets have been processed and come in the form of a matrix, $\bm{X} \in \R^{n \times p}$, where $n$ is the number of samples and $p$ is the number of gene expression measurements. We work with the $p=22{,}283$ genes shared by all datasets. The sample sizes vary (Mainz, $n_s=200$, Transbig, $n_s=198$, Vdx, $n_s=344$).
 
 \par Breast tumors can be classified into subtypes characterized by distinct molecular markers and clinical characteristics. There are four established molecular subtypes of breast cancer. 
 Luminal A tumors are less aggressive than the other subtypes, have lower proliferation,  express hormone receptors Estrogen Receptor (ER) and Progesteron Receptor (PR) and do not express the ERBB2 gene. 
These tumors are sensitive to endocrine therapy and have better prognosis compared  to  the  other subtypes. 
Luminal B tumors are ER and PR positive, have higher proliferation and may or may not have ERBB2 expression.
These tumors are less responsive to hormonal therapy and are more aggressive. The Her2+ subtype is characterized by increased expression of ERBB2  and low levels of ER and PR expression. This tumor subtype is  aggressive and patients do not respond to endocrine therapy. Treatment options include ERBB2-targeted therapies. Basal-like breast cancer subtype is characterized by low levels of ER, PR and ERBB2 expression. This is the most aggressive subtype with first and later therapy lines often limited to single-agent chemotherapy \citep{waks2019breast}. These subtypes were first identified by applying unsupervised hierarchical clustering to gene expression data from breast tumor tissues \citep{perou2000molecular}. In our experiments, for the purpose of comparisons, we take advantage of previously proposed subsets of genes: the PAM50 subset \citep{pam_50}, which contains fifty genes, and the three genes [3G] subset proposed by \citet{haibe2012three}. 
 
\subsection{Global replicability scores}

\begin{figure*}
         \includegraphics[width=\textwidth]{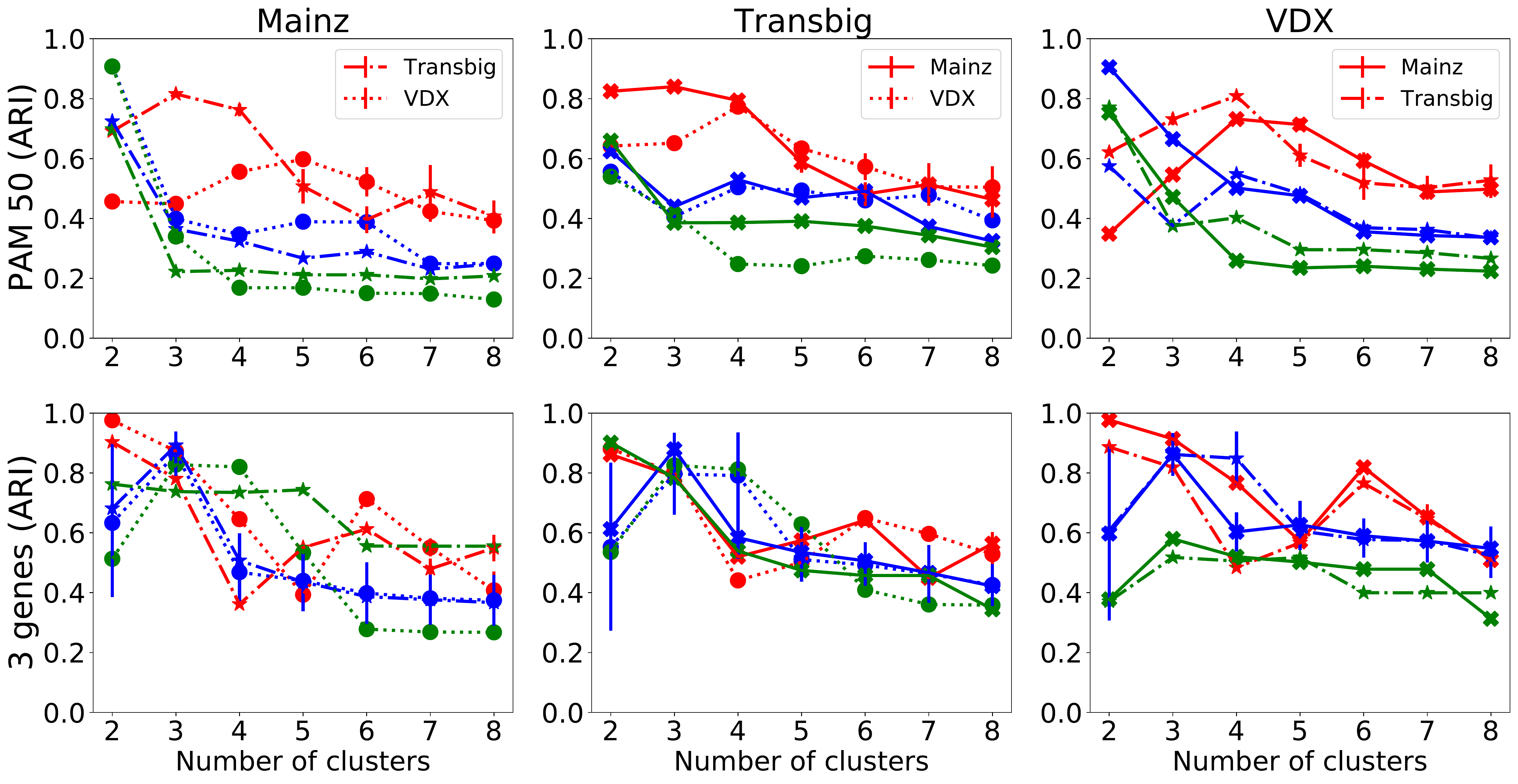} 
\caption{Average replicability scores using \Cref{algo_1} on cancer datasets of \citet{mainz,  transbig, vdx}. Each row identifies a gene subset (top: PAM50 \citep{pam_50}, bottom: 3G \citep{haibe2012three}). Each column identifies a testing datasets (from left to right: \citet{mainz, transbig, vdx}). For each pair of gene subset and test dataset, we separately consider each of the remaining two datasets as the training set. Line styles identify the training set used. We use $k$-means (red), Birch (green), and agglomerative clustering (blue),  for $k \in \{2,3,\dots,8\}$, and report the average replicability scores (ARI) across $B = 100$ iterations $\pm$ one standard deviation (vertical axis). }
\label{fig:real_1}
\end{figure*}

First, we compare the replicability scores of simple clustering methods using \Cref{algo_1} on the three datasets considered. We use in turn each pair of datasets as training and testing, and analyze how the replicability properties change as we vary (i) the replicability metric, (ii) the training/testing pair, (iii) the clustering algorithm and the number of clusters $k$, and (iv) the genes subsets (\Cref{fig:real_1}). Four main observations emerge: 
\begin{itemize}
\item The relative ranking of clustering methods with respect to replicability scores is robust to the choice of ARI or AMI.
\item Replicability scores are generally comparable across most pairs of training and testing datasets and produce similar results when the roles of $\bm{X}$ and $\bm{X}'$ are reversed.  
\item The choices of clustering algorithm and $k$ play a crucial role in determining replicability scores. 
\item When using the 3G subset, $k=3$ yields the highest replicability scores, while with the PAM 50 subset $k=4$  achieves the highest scores.
\end{itemize}  Our replicability metric provides consistent results with the subtypes published in \citet{pam_50} and \citet{haibe2012three} (see Figure 9 in \citet{masoero2022clusteringsupp}). Namely, clustering functions achieving high replicability scores produce partitions similar to the classification-based subtypes. This provides evidence that reproducible partitions are also consistent with well accepted biological findings.

\subsection{Local replicability scores}

Next, we perform a local replicability analysis using \Cref{algo_3} on the three datasets considered, using both the 3G and PAM50 gene subsets. We find that the cluster associated with the Basal subtype is the easiest to identify and the most robust, with the highest local replicability score. Luminal A and Luminal B are  harder to distinguish, and are associated with lower local replicability scores. 

\begin{figure*}
         \includegraphics[width=.8\textwidth]{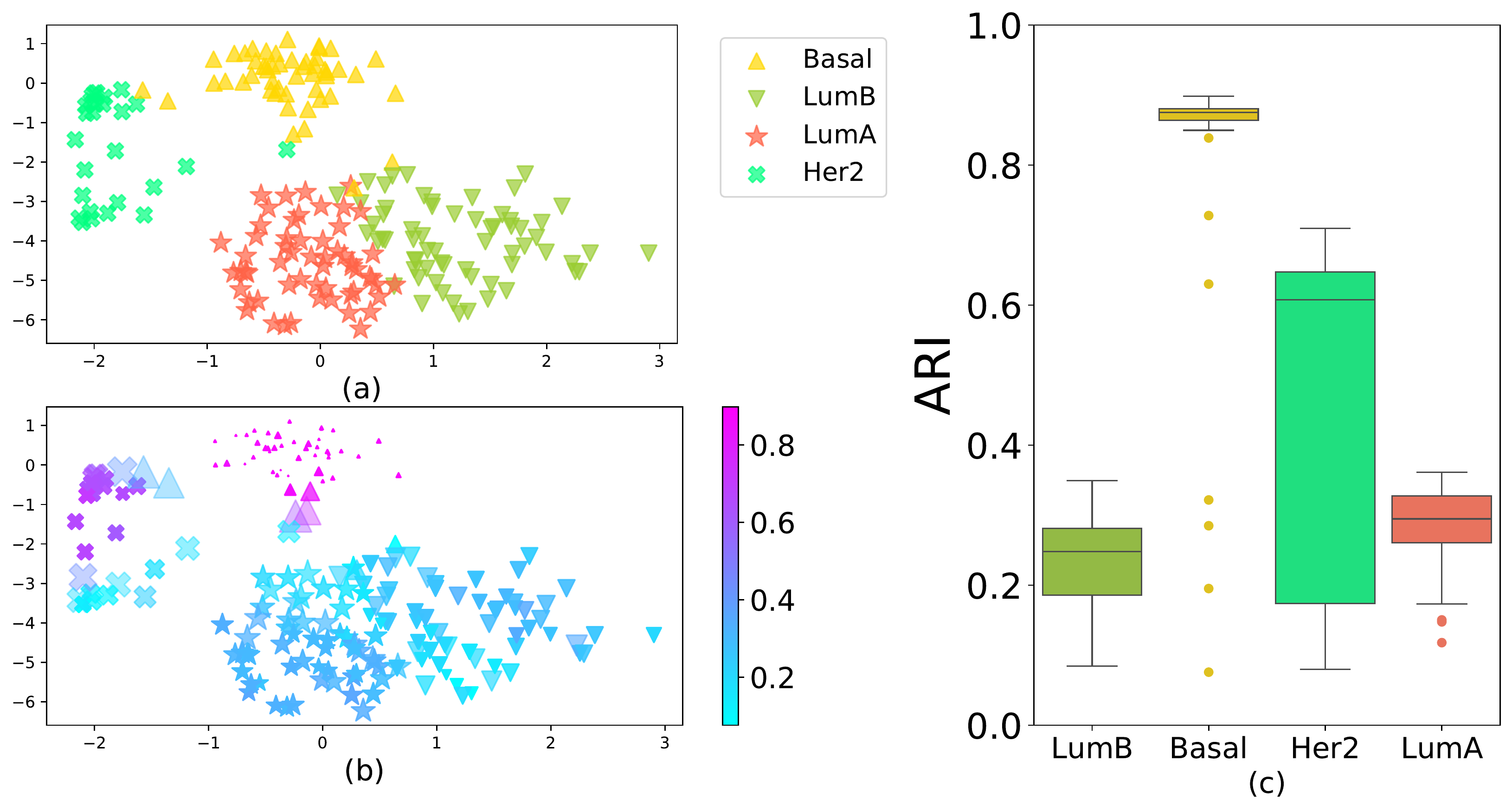} 
\caption{Local replicability of observed points in Transbig ($\bm{X}'$, \citet{transbig}) when
   Mainz  ($\bm{X}$, \citet{mainz}) is  used for training.
    We use the 3G subset \citep{haibe2012three} and $k$-means,  $k=4$. 
    In  (a), we visualize $\bm{X}'$. 
    Points are grouped into different clusters  according to $\Psi(\Xte;\Xtr)$, which  closely resembles $\Pi_{\texttt{3G}}(\Xte)$, so we can assign one cancer subtype to each cluster. 
    In  (b), we report the local replicability scores obtained by running \Cref{algo_3} on every point $x' \in \Xte$. 
    Colors reflects the average local replicability (AMI) score, while the size and transparency are proportional to the standard deviation across the bootstrap iterations: larger  points have more variable scores.
    We plot the data using a tSNE embedding with well separated clusters \citep{maaten2008visualizing}. 
 In  (c), for each cluster subtype (as defined by $\Psi(\Xte;\Xtr)$), we report ARI scores.}
\label{fig:local_repro_real_all}
\end{figure*}

In \Cref{fig:local_repro_real_all} we provide a visualization of our findings. We use the 3G subset, the Transbig ($\Xte$) dataset \citep{transbig} for testing, and the Mainz ($\Xtr$) dataset \citep{mainz} for training. To produce \Cref{fig:local_repro_real_all}, for each point $x \in \Xte$ we compute the local replicability score using \Cref{algo_3}, using $k$-means with $k=4$, $\Xtr$ as training set and $B=100$ bootstrap iterations. 
To group datapoints in $\Xte$ into different clusters, we use $\Psi(\Xte;\Xtr)$. The partition obtained closely resembles the model-based signature $\Pi_{\texttt{3G}}(\Xte)$ provided by \citet{haibe2012three}. We therefore match each learned cluster to one of the cancer subtypes labels, so that each cluster corresponded to one cancer subtype (Luminal A, Luminal B, Basal, Her2+). The average local replicability scores within each block of the partition confirm what expected: the Basal subtype is the most reproducible, while Luminal A and Luminal B are the least reproducible. 

%% file: sec-disc.tex
% !TEX root = main.tex

%%%%%%%%%%%%%%%%%%%%%%%%
\section{Discussion }
%%%%%%%%%%%%%%%%%%%%%%%%

In this paper, we provide a cohesive review of existing methods for replicability of clustering analyses, and develop a novel framework for replicability of clustering when multiple datasets are available. 
This new  approach  allows to quantify replicability with  any number of datasets and using any clustering algorithm,  at a local as well as at a global scale. In  our experiments, we show that our replicability scores guide the choice of an effective clustering algorithm and the tuning of relevant parameters, such as the number of clusters used in the analysis. Our evaluation procedures build on the  bootstrap method: using  bootstrap subsamples allows to quantify uncertainty summaries and interval estimates,  making  the replicability scores more informative (see \citet[Appendix A.1]{masoero2022clusteringsupp}). The bootstrap approach mitigates the impact of outliers  on the replicability scores. We report additional experiments  in \citet[Appendices A, C]{masoero2022clusteringsupp}: these are conducted using our method, as well as the methods reviewed in \Cref{sec:review}. Our experimental findings suggest that our method is a valuable tool for replicability analyses. In applications, our newly proposed metrics can help choosing which clustering algorithm to use, suggesting which method is best for the analysis of the data at hand. Last, while we have here focused on clustering algorithms, we emphasize that there exists a large literature on random partition models and statistical modeling that can be useful for clustering problems (see, e.g. \citet{muller2010random,wade2018bayesian}).

%% file: sec-acks.tex
\paragraph*{Acknowledgments} LT has been supported by the NIH grant 5R01LM013352-02 and the NSF grant 2113707.

%% file: sec-add.tex
%% !TEX root = appendix.tex
\section{Additional experiments on synthetic data} \label{sec:exp_add}

\subsection{Calibration checks for the bootstrap intervals: experimental setup} \label{sec:app_calib}

To assess the usefulness of the bootstrap intervals that are obtained as a byproduct of the output of Algorithm 1 in the main text \citep{masoero2022clustering}, we here present a procedure to measure the calibration of these intervals. For this purpose, let $F_1$ and $F_2$ be two fixed probability distributions with support on $\R^p$. Let $\AA$ be a fixed clustering algorithm.
\newline
\par Draw $n$ i.i.d.\ random replicates $x_1,\ldots, x_n$ from $F_1$, and let $\Xtr = [x_1,\ldots,x_n]^\top \in \R^{n\times p}$ be the training set. Similarly, draw $m$ i.i.d.\ random replicates $x'_1,\ldots, x'_m$ from $F_2$, and let $\Xte  = [x_1',\ldots,x_m']^\top \in \R^{m \times p}$ be the testing dataset. Run Algorithm 1, as detailed in the main text \citep{masoero2022clustering}, over a large number $B$ of bootstrap draws from $\Xtr$ and $\Xte$ respectively. Given the output $\hat{R}^{(1)}, \ldots, \hat{R}^{(B)}$, and for a fixed $\alpha \in [0,1]$, let $\hat{R}_{(\alpha)}$ be the $100\times\alpha \%$  quantile of the bootstrap scores $\hat{R}^{(1)}, \ldots, \hat{R}^{(B)}$. 
\newline 
\par To assess the quality of the bootstrap intervals obtained from $\hat{R}_{(\alpha)}$, we compare them to corresponding Monte Carlo values, obtained by repeatedly drawing from $F_1$ and $F_2$. In detail, fix a large $N_{MC}$. For each $s =1 ,\ldots, N_{MC}$, draw $n$ i.i.d.\ random replicates $z^{(s)}_1,\ldots, z^{(s)}_n$ from $F_1$, and $m$ i.i.d.\ random replicates $w^{(s)}_1,\ldots, w^{(s)}_m$ from $F_2$; let $\bm{Z}^{(s)} = [z^{(s)}_1,\ldots, z^{(s)}_n]^\top$ and $\bm{Z}^{'(s)} = [w^{(s)}_1,\ldots, w^{(s)}_m]^\top$. Compute the replicability score $\hat{\hat{R}}^{(s)} := r^\star(\Psi(\bm{Z}^{'(s)} ; \AA, \bm{Z}^{(s)} ), \Psi(\bm{Z}^{'(s)} ;\AA, \bm{Z}^{'(s)} ))$. Given the output $\hat{\hat{R}}^{(1)}, \ldots, \hat{\hat{R}}^{(N_{MC})}$, and for a fixed $\alpha \in [0,1]$, let $\hat{\hat{R}}_{(\alpha)}$ be the $100\times\alpha \%$ quantile of the Monte Carlo scores $\hat{\hat{R}}^{(1)}, \ldots, \hat{\hat{R}}^{(B)}$. We provide a visual diagnostic, in which we compare the extent to which the Monte Carlo and Bootstrap intervals are similar or differ.

\subsection*{Experimental details}
We present experimental results for the setup described above. Here, we let $n=m =500$, $p=2$, $B=1000$, and $N_{MC}=1000$. The algorithm $\AA$ is $k$-means. We also let $F_1$, $F_2$ be mixtures of Gaussians, of the form:
\[
	F_1 = \frac{1}{4}\sum_{j=1}^4\mathcal{N}(\mu_j, \sigma I), \quad F_2 = \frac{1}{4}\sum_{j=1}^4\mathcal{N}(\mu'_j, \sigma I),
\]
with $\sigma =1$, and $\mu_1 = [0,-7], \mu_2 = [3.5, 3], \mu_3 = [-2,2], \mu_4 = [2,-2]$ and $\mu'_1 = [-1,-7], \mu_2 = [4.2, 3.3], \mu_3 = [-2.5,1.8], \mu_4 = [2.2,-3]$.

\begin{figure}[h!]
\centering\includegraphics[width = .9\textwidth]{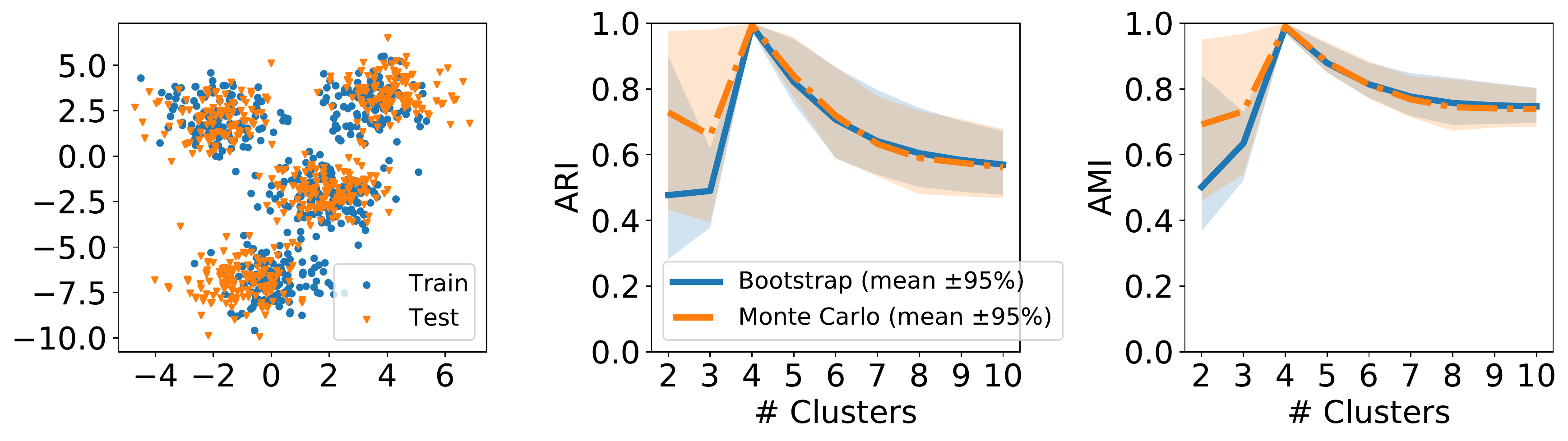}
    \caption{Left: draws $\Xtr \in \R^{500,2}$ (blue), $\Xte \in \R^{500,2}$ (orange) from train and test distributions $F_1$ and $F_2$ respectively. Center: ARI replicability score when using $k$-means (vertical axis) for different choices of $k$ (horizontal axis). The blue line reports the average score over $B=1000$ bootstrap replicates for the two datasets plotted in the left subplot. The orange dotted line reports the average score over $N_{MC} = 1000$ Monte Carlo draws of different pairs of training and testing datasets, drawn the same distributions $F_1$, $F_2$ above. Shadowed regions cover a centered $95\%$ interval for the two scores. }
    \label{fig:caibration_1}
\end{figure}

We compare the average replicability score over the Monte Carlo and the bootstrap replicates, together with a centered $95\%$ interval, as we vary the choice $k$ of the number of clusters in \Cref{fig:caibration_1}. We find that, across $k$, the bootstrap and Monte Carlo values are very close, showing that the bootstrap intervals enjoy good calibration.

\begin{figure}[h!]
\centering \includegraphics[width = \textwidth]{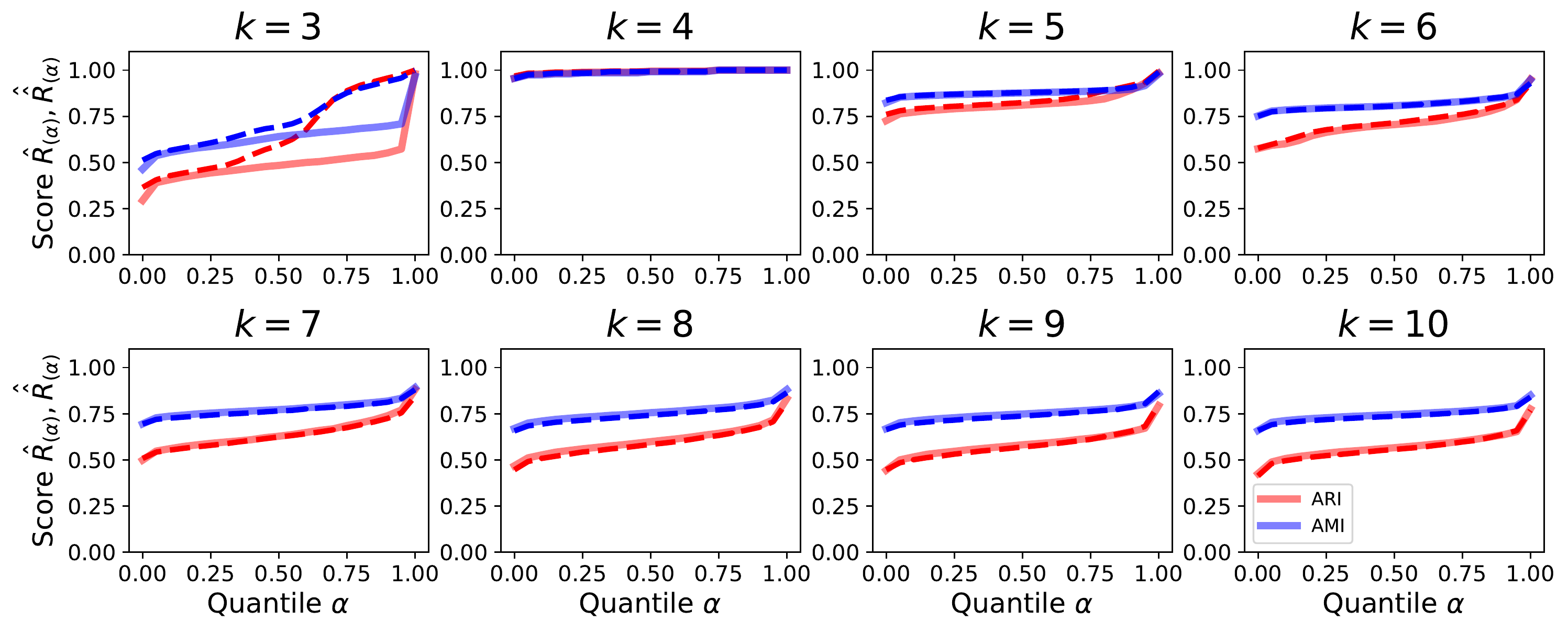}
    \caption{Quantiles coverage. For the same distributions $F_1$, $F_2$ considered in \Cref{fig:caibration_1}, each subplot reports the bootstrap (solid lines, $\hat{R}_{(\alpha)}$) and Monte Carlo (dotted lines, $\hat{\hat{R}}_{(\alpha)}$) score (vertical axis) obtained as we increase the quantile (horizontal axis). Red lines refer to ARI, blue lines to AMI.} \label{fig:caibration_quant_1}
   \end{figure}

Next, we inspect in \Cref{fig:caibration_quant_1} the calibration of the bootstrap quantiles with respect to the Monte Carlo quantiles. That is, for each $\alpha$ in $[0,1]$, we compare the values $\hat{R}_{(\alpha)}$ and $\hat{\hat{R}}_{(\alpha)}$. We perform this comparison across different values of $k$, and verify that the bootstrap and Monte Carlo quantiles are close.
   
\begin{figure}[h]
\includegraphics[width = \textwidth]{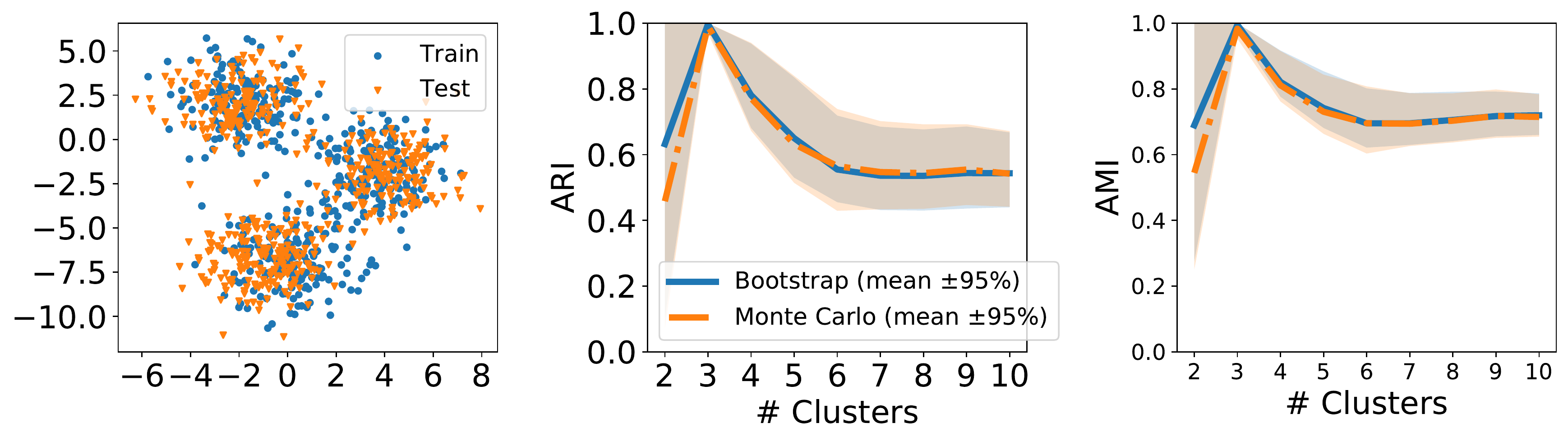}
    \caption{We use the same setup as \Cref{fig:caibration_1}, now for data drawn from the distributions in \Cref{eq:other_dist}.}\label{fig:calibration_2}
\end{figure}

\begin{figure}[h]
\includegraphics[width = \textwidth]{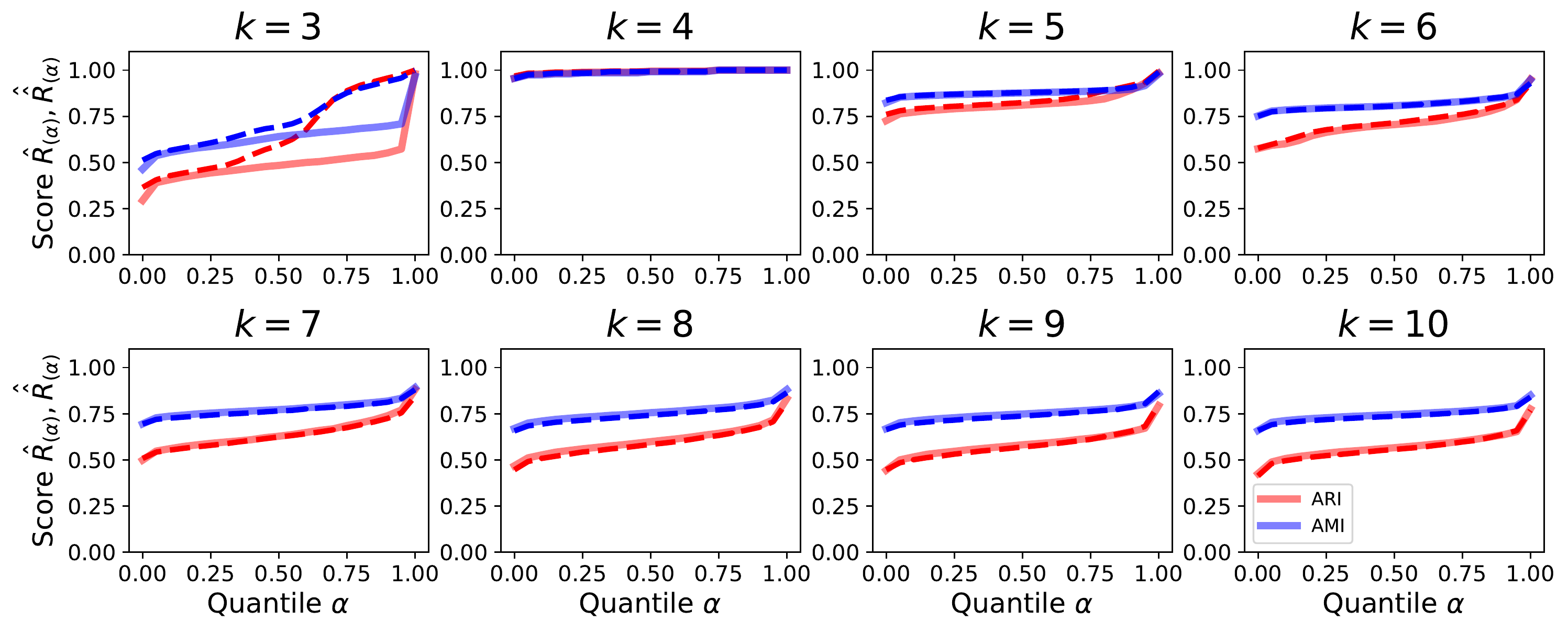}
    \caption{We use the same setup as \Cref{fig:caibration_quant_1}, now for data drawn from the distributions in \Cref{eq:other_dist}.} \label{fig:caibration_quant_2}
   \end{figure}

We repeat the same experiments, for different choices of $F_1$ and $F_2$, as defined below:
\begin{equation}
	F_1 = \frac{1}{4}\sum_{j=1}^4\mathcal{N}(\mu_j, \sigma I), \quad F_2 = \frac{1}{4}\sum_{j=1}^4\mathcal{N}(\mu'_j, \sigma I), \label{eq:other_dist}
\end{equation}
now with $\mu_1 = [0,-7], \mu_2 = [3.5, -2], \mu_3 = [2,-2]$, and $\mu_1' = [-1,-7], \mu_2' = [4.2, -1.8], \mu_3' = [-2.5, 1.8]$. Plots are included in \Cref{fig:calibration_2,fig:caibration_quant_2}.

\subsection{Replicability scores in the absence of clustering structure}
To test whether our replicability metric is robust to the absence of clustering structure, we perform the following experiment. For fixed $n$, we draw $2n$ i.i.d.\ replicates $x_1, \ldots, x_{2n}$ from a fixed distribution $F$. We let $\Xtr = [x_1,\ldots,x_n]^\top$ and $\Xte = [x_{n+1},\ldots,x_{2n}]^\top$ be the training and testing set respectively, and apply Algorithm 1 to inspect the behavior of the replicability scores on this data.

\begin{figure}[h!]
\centering \includegraphics[width = \textwidth]{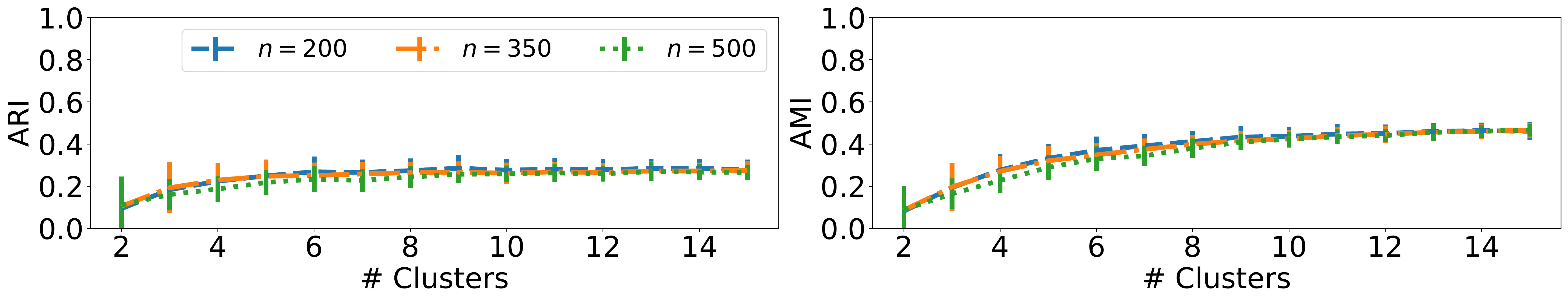}
    \caption{Replicability scores in the absence of clustering structure. We report the average replicability score $\pm$ one standard deviation as obtained by $B=100$ replicates according to Algorithm 1 (vertical axis, left: ARI, right: AMI) as we vary the number of clusters using $k$-means. Different line colors and styles refer to couples of training, testing datasets with different sizes $n$. Datapoints in the train and test dataset are i.i.d.\ multivariate Gaussian random draws with mean $0$ and identity covariance matrix in $\R^2$.} \label{fig:no_clustering}
   \end{figure}

In our experiments below, we let $F$ be the multivariate ``standard'' Gaussian, with mean the zero vector and covariance matrix the identity. In \Cref{fig:no_clustering} we fix $d = 2$ and vary $n\in\{200,350,500\}$. The qualitative finding from both the ARI and AMI scores are reassuring: in both cases the scores are low, and uniform across different choices of $k$.
   
Next, we inspect the behavior of the replicability scores across different choices of the dimension $d$ of the random vectors drawn --- $d \in \{5,10,15\}$, to understand the role of dimensionality. Again, scores are low, and uniform across different choices of $k$. Here $n=200$.

   \begin{figure}[h!]
\centering \includegraphics[width = \textwidth]{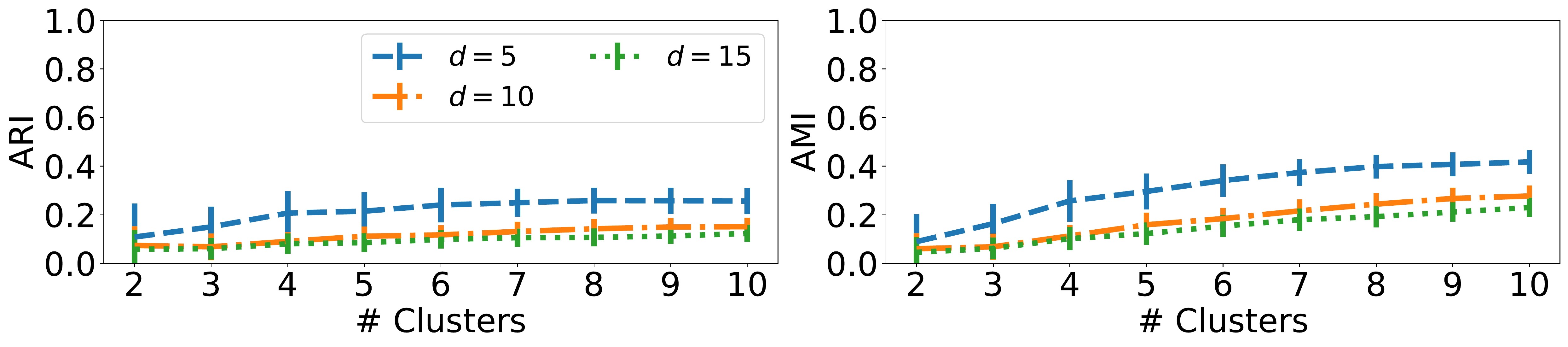}
    \caption{Replicability scores in the absence of clustering structure. We report the average replicability score $\pm$ one standard deviation as obtained by $B=100$ replicates according to Algorithm 1 (vertical axis, left: ARI, right: AMI) as we vary the number of clusters using $k$-means. Different line colors and styles refer to couples of training, testing datasets with different dimension $d$. Datapoints in the train and test dataset are i.i.d.\ multivariate Gaussian random draws with mean $0$ and identity covariance matrix in $\R^d$. Here, $n=200$.} \label{fig:no_clustering_d}
   \end{figure}

\subsection{The role of sample size in replicability metrics} \label{sec:sample_sizes}

\begin{figure}[h!]
\centering \includegraphics[width = \textwidth]{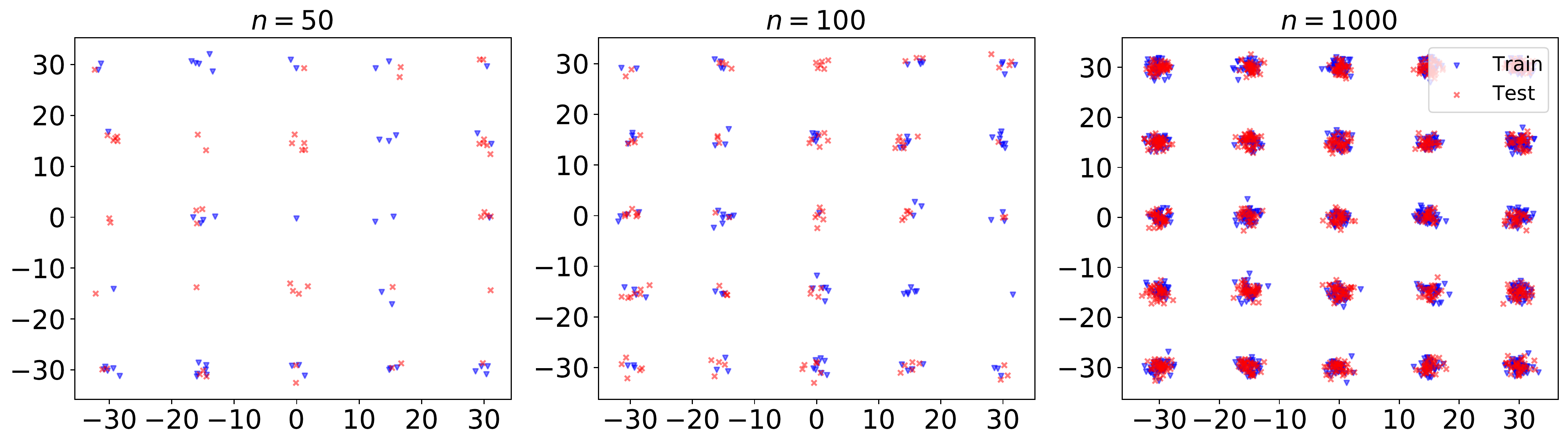}
    \caption{Three draws from a Gaussian mixture model in $\R^2$ with $k=25$ well separated components. Left: $n=50$, center: $n=100$, right: $n=1000$.} \label{fig:n_dependence_ds}
 \end{figure}

\begin{figure}[h!]
\centering \includegraphics[width = \textwidth]{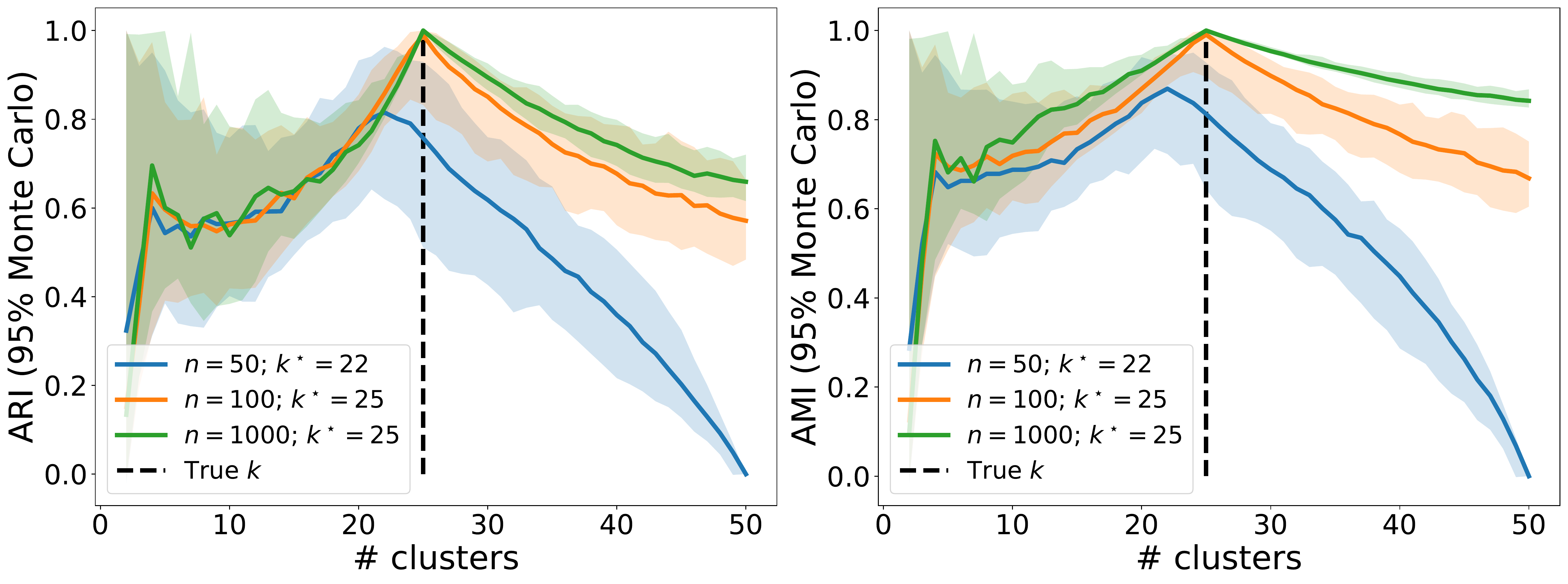}
    \caption{ARI and AMI replicability scores (vertical axis) across $N_{MC} = 100$ (Monte Carlo replicates) draws, for $n \in \{50,100,1000\}$, as the input parameter $k$ of $k$-means varies (horizontal axis). For each $n$, we report the mean (solid line) when $m=n$, as well as centered $95\%$ intervals across $k \in \{2,3,\ldots,50\}$. } \label{fig:n_dependence}
 \end{figure}

The replicability measures defined in Section 3 depend on the sample sizes of the training and testing datasets. A simple example to explain why in our  opinion this  is appropriate, is as follows: consider a training dataset $\Xtr$ with $n$ observations drawn i.i.d.\ from $F_1$, and a testing dataset $\Xte$ with $m$ observations drawn i.i.d.\ from $F_2$. For simplicity, let $n=m$, and $F_1,F_2$ be the same distribution (call it $F$), e.g.\  a mixture of Gaussians with a large number of well-separated components, say $k=25$, with equal mixing weights over the components. When $n$ is very small, e.g. $n=50$, with high probability some of the components of $F$ will be present in the training set, but absent in the testing set, and viceversa. We expect the replicability index to capture this phenomenon --- in this case, for example, there might exist a value $k<25$ which achieves a higher replicability score than $k=25$. As $n$ increases, however, with high probability both $\Xtr$ and $\Xte$ will contain datapoints from all components, and $k=25$ will achieve the highest replicability score. We include a brief simulation experiment below --- see \Cref{fig:n_dependence_ds,fig:n_dependence}.
\newline
\par We emphasize that in applications with  limited sample  sizes, anticipating trends like the one discussed above is challenging, and it is  not  part of  the  aims of our  manuscript.  The  focus  is  on replicability  of  the  cluster analysis results  with  the  available sample sizes, and not  with  hypothetical large  sample sizes. 

\section{Additional figures and  information about model-based signatures} \label{sec:additional-figures}

 \begin{center}
\begin{figure}[H]
         \includegraphics[width=\textwidth]{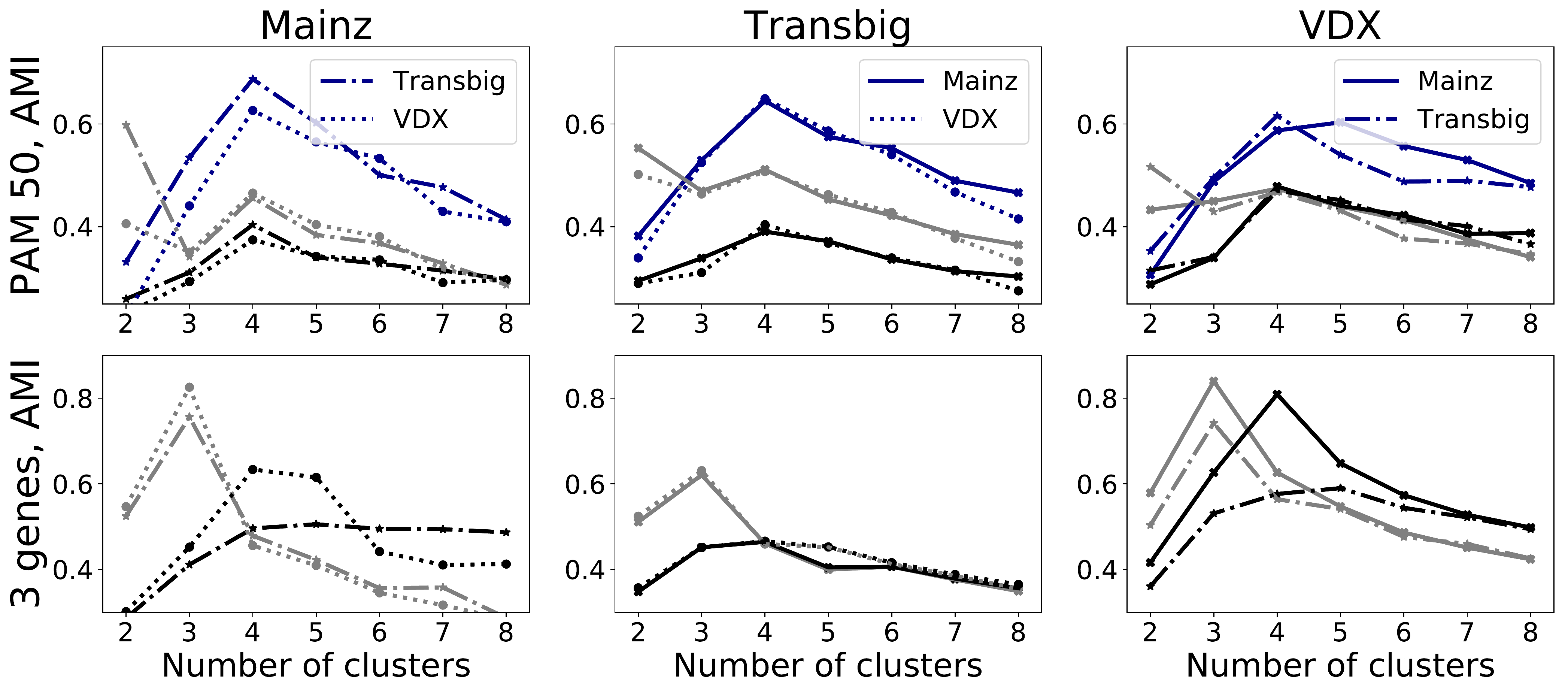} 
\caption{Each row refers to a different subset of genes (top row: PAM50 of \citet{pam_50}, bottom row: 3G of \citet{haibe2012three}). Each column refers to a different testing dataset $\bm{X}'$ (left: Mainz, center: Transbig, right: VDX).
For each gene subset, given testing dataset $\bm{X}'$, training dataset $\bm{X}$, algorithm $\AA$ and number of clusters $k$, we retain the partition $\Psi^*_{\bm{X}}(\bm{X}')$ with the highest replicability score across replications and compare it to known signatures $\Pi_{\texttt{PAM50}}(\bm{X}')$ (top row) and $\Pi_{\texttt{3G}}(\bm{X}')$ (bottom row) by computing $\AMI(\Psi^*_{\bm{X}}(\bm{X}'), \Pi_{\texttt{PAM50}}(\bm{X}'))$ (top row  -- $y$-axis) and $\AMI(\Psi^*_{\bm{X}}(\bm{X}'), \Pi_{\texttt{3G}}(\bm{X}'))$ (bottom row --  $y$-axis).}
\label{fig:gold_std}
\end{figure}
\end{center}

\begin{table}
\centering
\begin{tabular}{|c|c|c|c|}
\hline
                       & \textbf{Mainz} & \textbf{Transbig} & \textbf{Vdx} \\ \hline
\textbf{$\ARI(\Pi_{\texttt{PAM50}}(\bm{X}),\Pi_{\texttt{3G}}(\bm{X}) )$}           & 0.33           & 0.34              & 0.53         \\ \hline
\textbf{$\AMI(\Pi_{\texttt{PAM50}}(\bm{X}),\Pi_{\texttt{3G}}(\bm{X}))$}           & 0.38           & 0.39              & 0.49         \\ \hline

\end{tabular}
\caption{Comparison of the subtype signatures for Mainz, Transbig and Vdx using either the three genes subset, or the PAM50 subset. For example, for the Mainz dataset, letting $\Pi_{\texttt{PAM50}}(\texttt{mainz})$ and $\Pi_{\texttt{3G}}(\texttt{mainz})$ be the PAM50 and 3G signatures (i.e.\, row-wise partitions) of the Mainz dataset, we compute
$\ARI(\Pi_{\texttt{PAM50}}(\texttt{mainz}),\Pi_{\texttt{3G}}( \texttt{mainz})) = 0.33$ and $\AMI(\Pi_{\texttt{PAM50}}(\texttt{mainz}),\Pi_{\texttt{3G}}(\texttt{mainz}))  = 0.38$.}
\label{tab:comparison}
\end{table}

%% file: sec-other_methods.tex
% !TEX root = appendix.tex

\section{Methods from previous literature} \label{sec:exp_other}

In this section we provide experimental results for the competing methods described in Section 2. All code to replicate our experiments and reproduce figures is available at \url{https://github.com/lorenzomasoero/clustering_replicability}.

%%%%%%%%%%%%%%%%%%%%%%%%
\subsection{Clustering replicability via stability} \label{sec:stability_exp}
%%%%%%%%%%%%%%%%%%%%%%%%

Here we report results for the methods discussed in Section 2.2.  We consider the datasets of \citet{ClusteringDatasets}.

%%%%%%%%%%%%%%%%%%%%%%%%
\subsubsection{Smolkin-Ghosh inclusion score}
%%%%%%%%%%%%%%%%%%%%%%%%

\begin{figure}
         \includegraphics[width=1\textwidth]{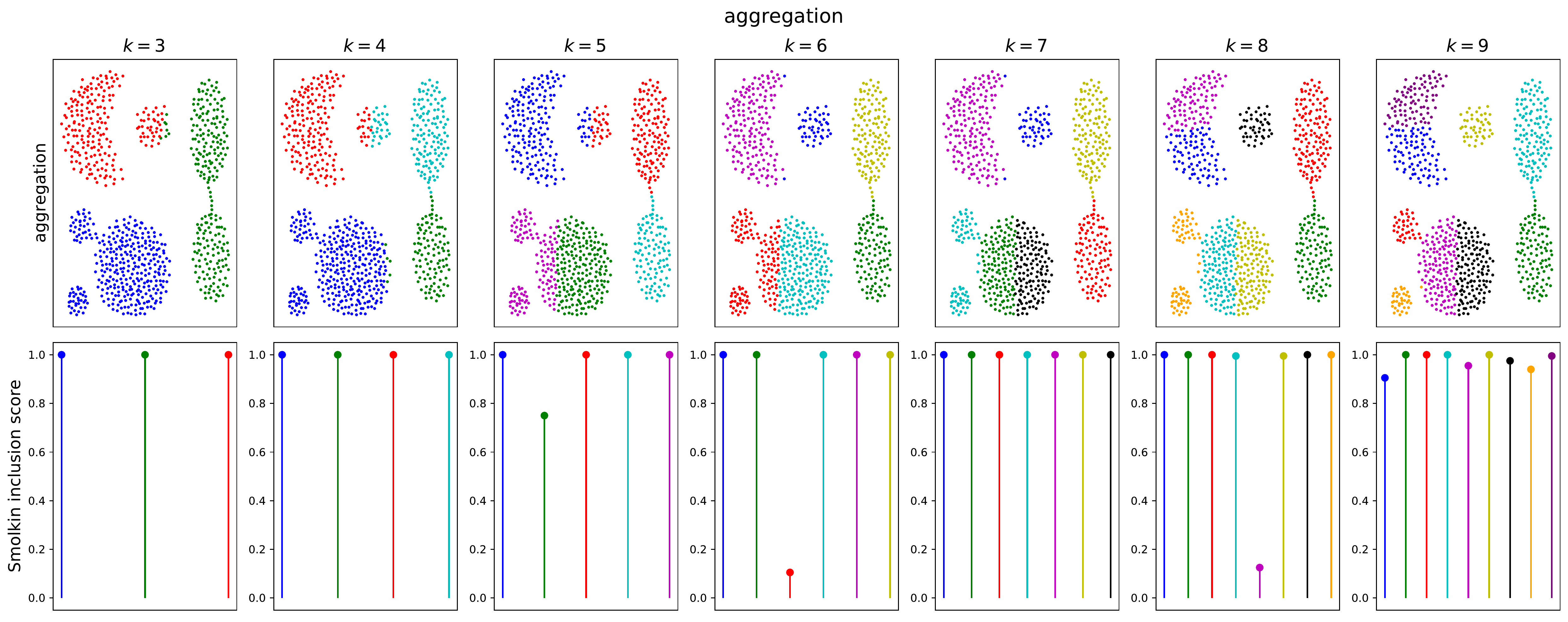} 
\caption{Smolkin-Ghosh inclusion score computed for the ``aggregation'' dataset using $k$-means, for different values of $k$.}
\label{fig:smolkin_aggregation}
\end{figure}

\begin{figure}
         \includegraphics[width=1\textwidth]{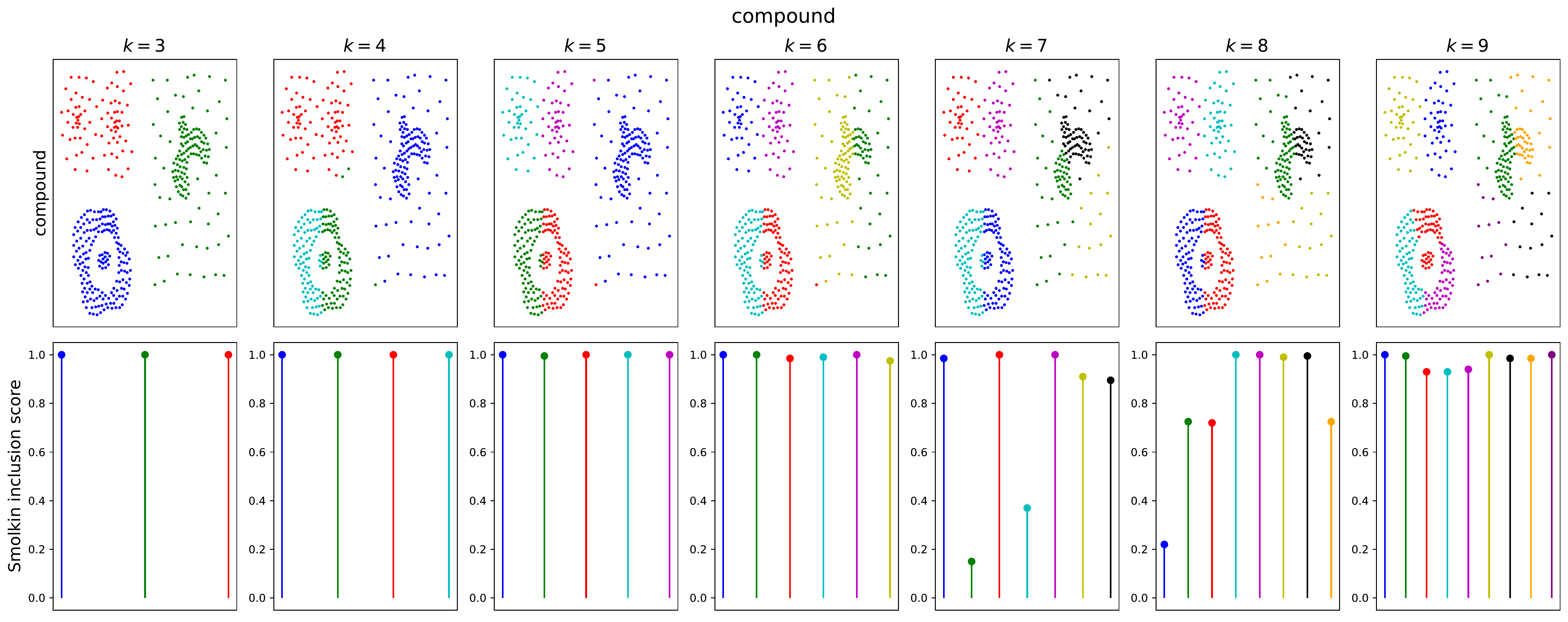} 
\caption{Smolkin-Ghosh inclusion score computed for the ``compound'' dataset using $k$-means, for different values of $k$.}
\label{fig:smolkin_compound}
\end{figure}

\begin{figure}
         \includegraphics[width=1\textwidth]{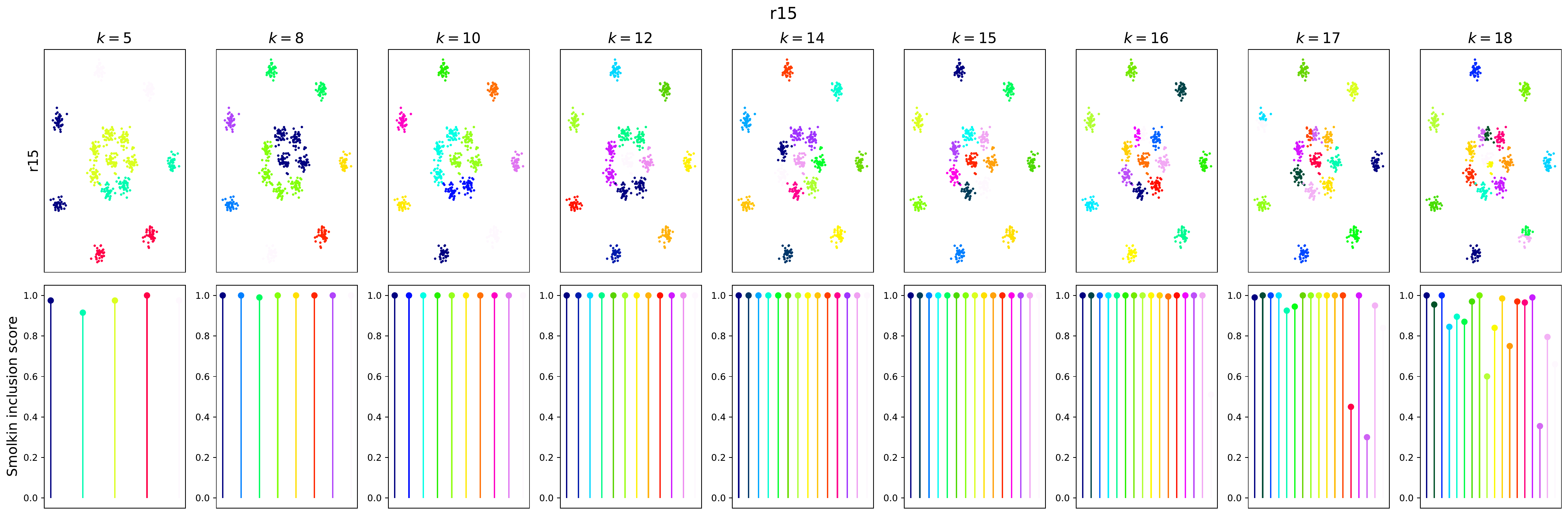} 
\caption{Smolkin-Ghosh inclusion score computed for the ``r15'' dataset using $k$-means, for different values of $k$.}
\label{fig:smolkin_r15}
\end{figure}

We start with the Smolkin-Ghosh inclusion score, proposed by \citet{smolkin2003cluster} and described in Section 2.2.2. Since the datasets we consider are not very high-dimensional, we let $\alpha = 1$ in our experiments (i.e., retain all the covariates for the purposes of clustering). All our results are obtined by averaging over $B=200$ bootstrap re-samples, using the $k$-means algorithm for different values of $k$. In \Cref{fig:smolkin_aggregation,fig:smolkin_compound,fig:smolkin_r15}, we color-code in the first row the original clusters $U_1,\ldots,U_k$ obtained by running $k$-means. In the second row, for each cluster, we report the corresponding Smolkin-Ghosh inclusion score.

In our experiments using the Smolkin-Ghosh in \Cref{fig:smolkin_aggregation}, we see that for $k=3,4,5$, the replicability scores of all the clusters are close to $1$ --- the highest possible value. However, by visually inspecting the clustering learned, we notice that some of the clusters learned are spurious.

%%%%%%%%%%%%%%%%%%%%%%%%
\subsubsection{D Index}
%%%%%%%%%%%%%%%%%%%%%%%%

\begin{figure}
         \includegraphics[width=1\textwidth]{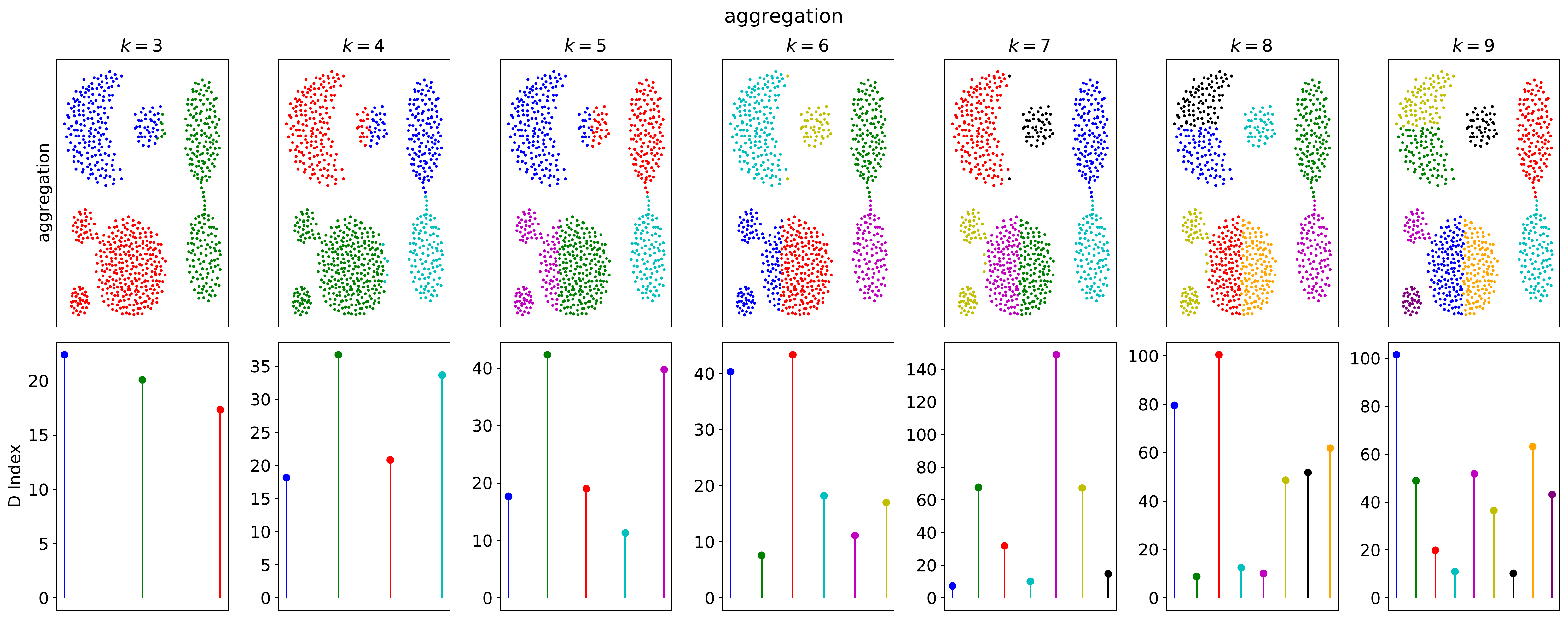} 
\caption{D index computed for the ``aggregation'' dataset using $k$-means, for different values of $k$.}
\label{fig:d_aggregation}
\end{figure}

\begin{figure}
         \includegraphics[width=1\textwidth]{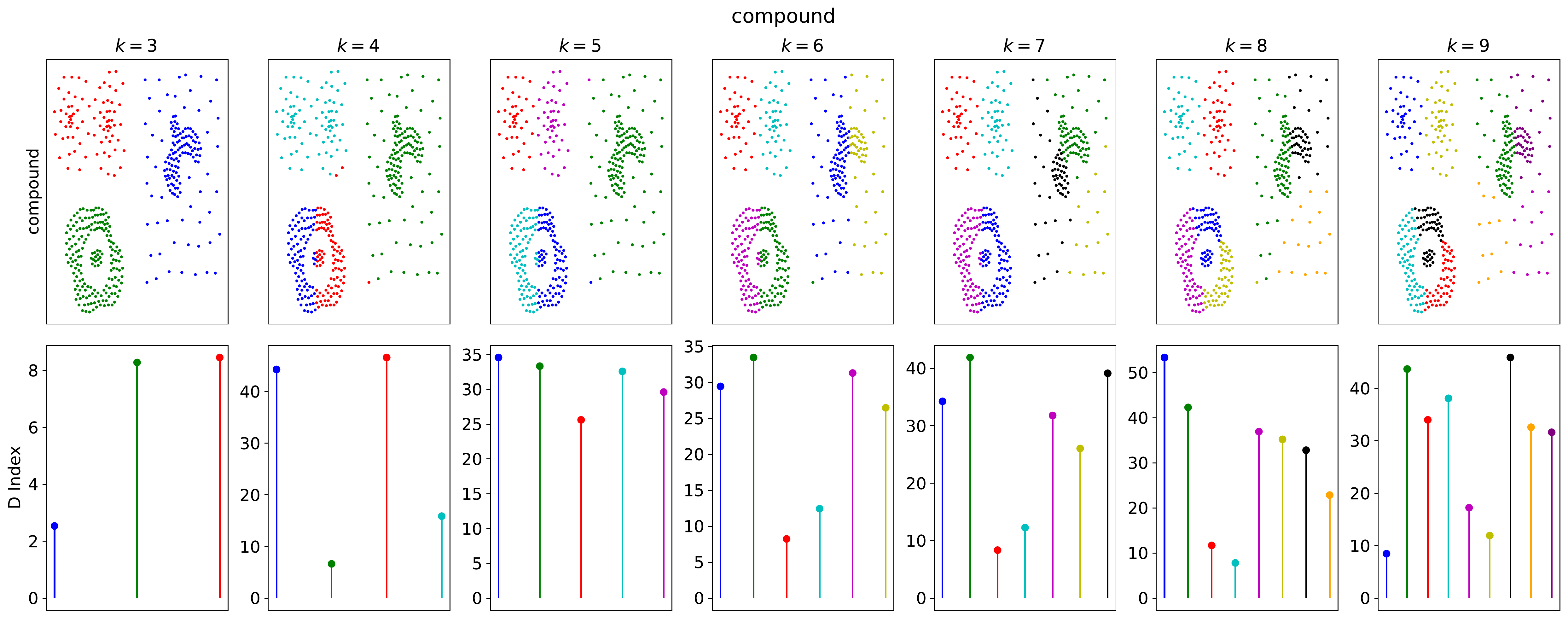} 
\caption{D index computed for the ``compound'' dataset using $k$-means, for different values of $k$.}
\label{fig:d_compound}
\end{figure}

\begin{figure}
         \includegraphics[width=1\textwidth]{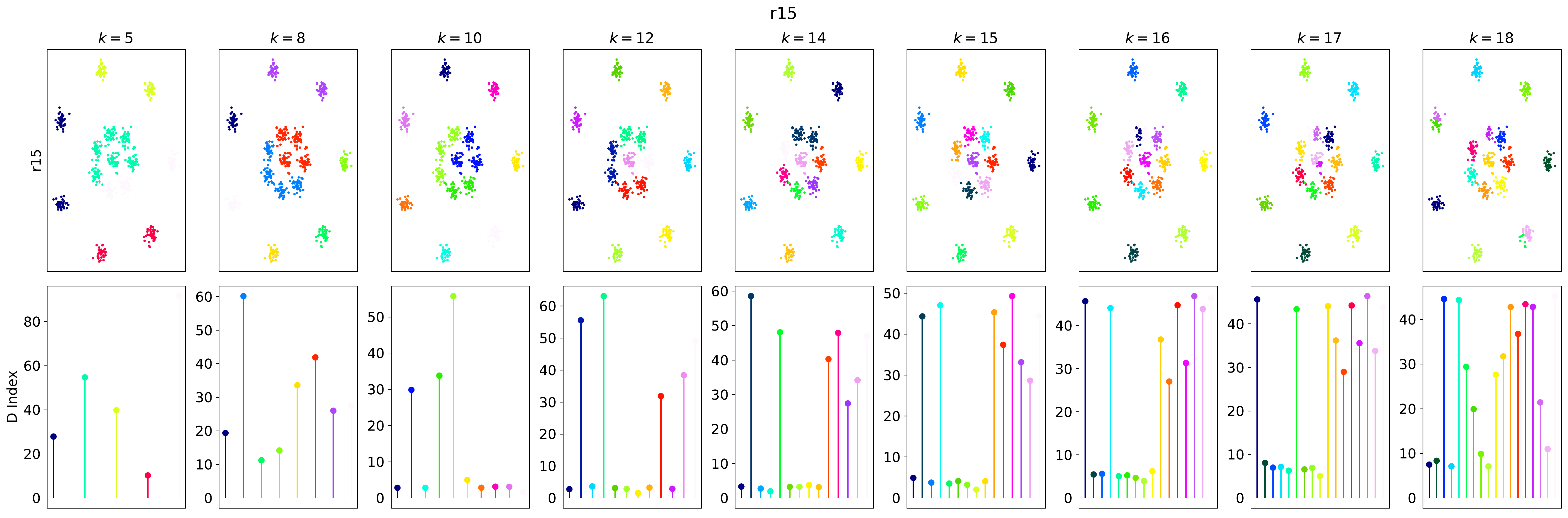} 
\caption{D index computed for the ``r15'' dataset using $k$-means, for different values of $k$.}
\label{fig:d_r15}
\end{figure}

Next we consider the deletion (D) index, proposed by \citet{mcshanerepro} and described in Section 2.2.2. In \Cref{fig:d_aggregation,fig:d_compound,fig:d_r15}, we color-code in the first row the original clusters $U_1,\ldots,U_k$ obtained by running $k$-means. In the second row, for each cluster, we report the corresponding D index score obtained by clustering $B=200$ perturbed versions of the original data. We perturb via additive Gaussian noise with variance given by $\hat{S}^2/2$, where $\hat{S}^2$ is the sample variance for the dataset under consideration (i.e., the $n$-th point in the $b$-th perturbed dataset is given by $\tilde{x}_n = x_n + e_n^{(b)}$, with $e_n^{(b)} \sim \mathcal{N}(0, \hat{S}^2)$).

In our experiments, we find that the D index can be useful to identify spurious clusters. One limitation of the score is that --- as is --- the D index provides an  (average) absolute number of deletions per cluster over the random re-samples. This can make the stability comparison between different clusters unfair, especially if their sizes are unbalanced. Therefore, we believe that reporting the average \emph{relative} number of deletions (with respect to the original cluster size) over the random re-samples allows for a more direct and effective comparison of the stability of clusters.

%%%%%%%%%%%%%%%%%%%%%%%%
\subsubsection{R Index}
%%%%%%%%%%%%%%%%%%%%%%%%

\begin{figure}
         \includegraphics[width=1\textwidth]{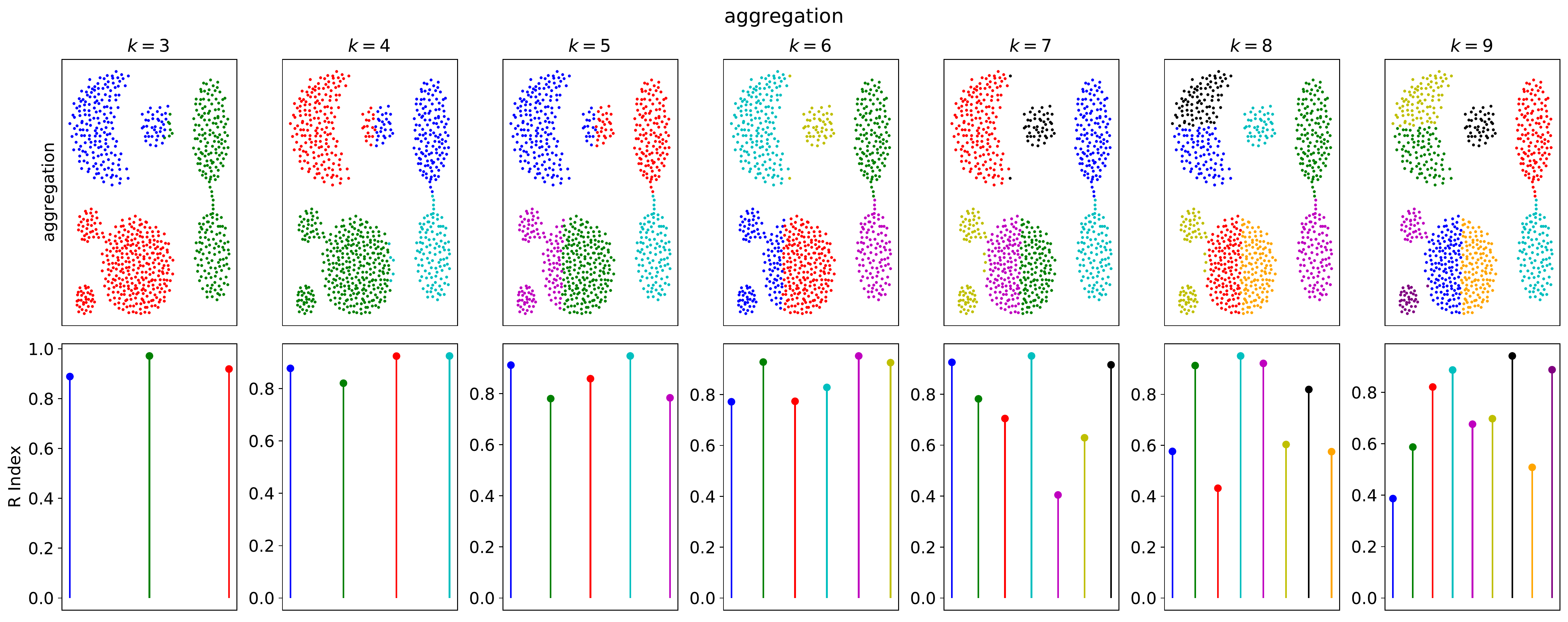} 
\caption{R index computed for the ``aggregation'' dataset using $k$-means, for different values of $k$.}
\label{fig:r_aggregation}
\end{figure}

\begin{figure}
         \includegraphics[width=1\textwidth]{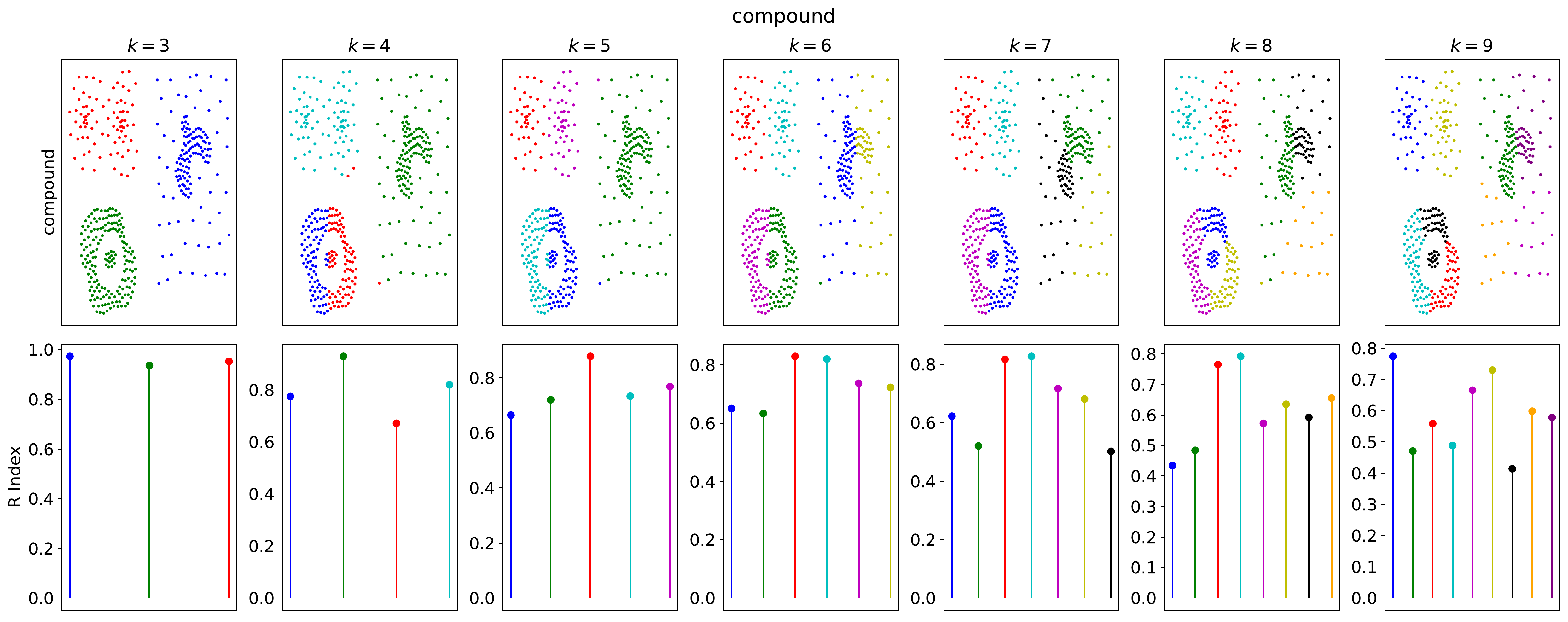} 
\caption{R index computed for the ``compound'' dataset using $k$-means, for different values of $k$.}
\label{fig:r_compound}
\end{figure}

\begin{figure}
         \includegraphics[width=1\textwidth]{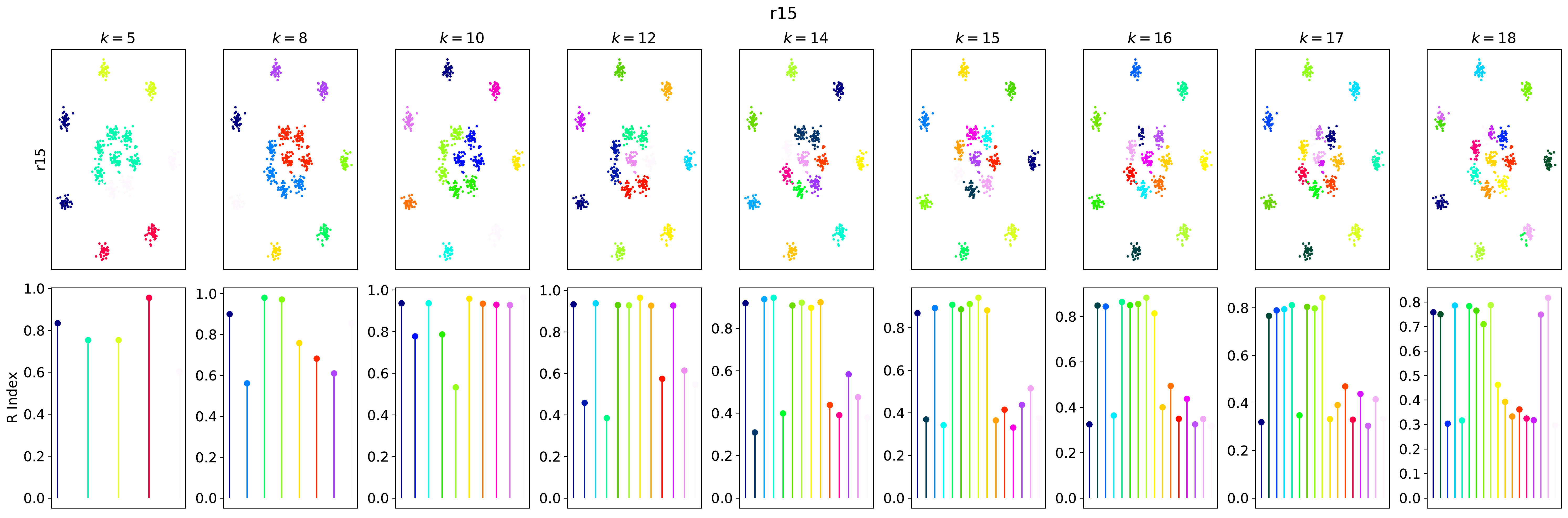} 
\caption{R index computed for the ``r15'' dataset using $k$-means, for different values of $k$.}
\label{fig:r_r15}
\end{figure}

Next we consider the R index, proposed by \citet{mcshanerepro} and described in Section 2.2.2. In \Cref{fig:r_aggregation,fig:r_compound,fig:r_r15}, we color-code in the first row the original clusters $U_1,\ldots,U_k$ obtained by running $k$-means. In the second row, for each cluster, we report the corresponding R index score obtained by clustering $B=200$ perturbed version of the original data. We perturb via additive Gaussian noise with variance given by $\hat{S}^2/2$, where $\hat{S}^2$ is the sample variance for the dataset under consideration (i.e., the $n$-th point in the $b$-th perturbed dataset is given by $\tilde{x}_n = x_n + e_n^{(b)}$, with $e_n^{(b)} \sim \mathcal{N}(0, \hat{S}^2)$).

In our experiments, we find that the R index can be a useful metrics to identify which cluster is more stable among the ones found. We also find that, in certain instances, the value of the score can be misleading. For example, in \Cref{fig:r_aggregation}, for $k=5$, while the algorithm fails to effectively identify well separated clusters, the R scores are not particularly low. This might lead to interpretability issues in high dimensional settings. As a side note, the findings of the R and D index heavily depend on the specification of the variance of the noise. We recommend running sensitivity analyses and carefully choosing this parameter when running analyses based on these indices.

%%%%%%%%%%%%%%%%%%%%%%%%
\subsection{Clustering replicability via prediction accuracy} \label{sec:prediction_exp}
%%%%%%%%%%%%%%%%%%%%%%%%

Here we report results for the methods discussed in Section 2.3. Again, we consider the datasets of \citet{ClusteringDatasets}.

\begin{figure}
         \includegraphics[width=1\textwidth]{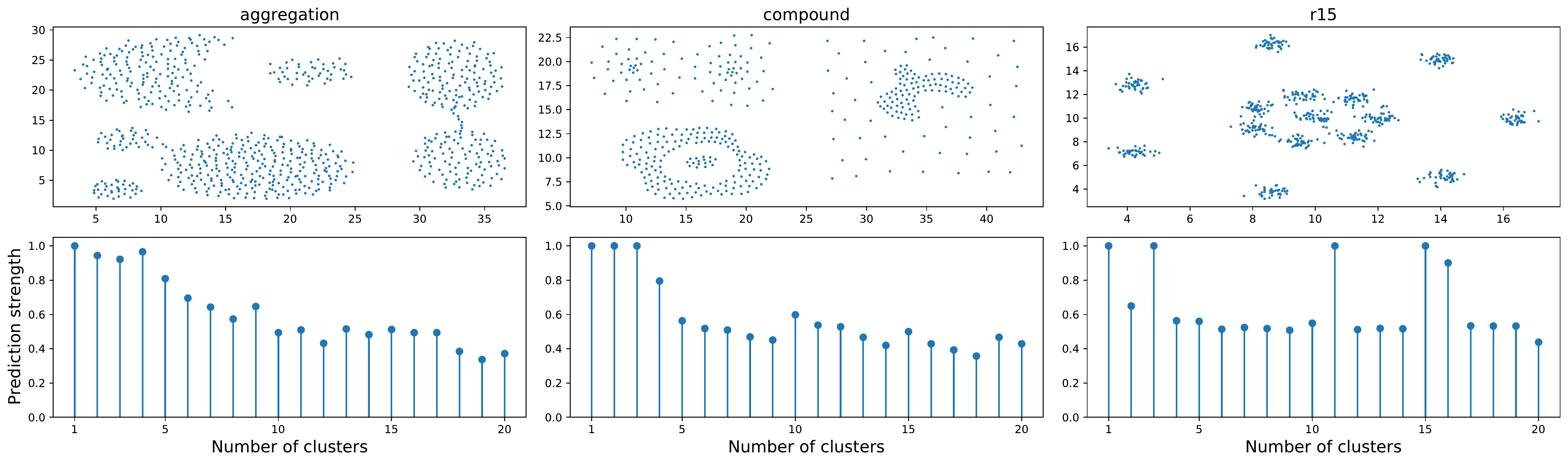} 
\caption{Prediction strength computed for the ``aggregation', ``compound'' and ``r15'' datasets using $k$-means, for different values of $k$.}
\label{fig:prediction_strength}
\end{figure}

We start by providing results for the prediction strength introduced by \citet{tibshirani2005cluster} for the three datasets already considered in the previous section. Results are reported in \Cref{fig:prediction_strength}. One of our findings --- consistent across our experiments --- is that the prediction strength tends to favor the choice of fewer clusters, even when such choice does not result in a better clustering. This is due to the fact that the prediction strength is obtained by applying the minimum operator across a cluster-specific score (see Equation (4) in the main manuscript \citep{masoero2022clustering}). Practitioners using the prediction score as a diagnostic tool to assess clustering replicability should be cautious and aware of this feature. We suggest monitoring all the cluster-specific scores that are contained inside the $\min$ operator in Equation (4)  in the main manuscript \citep{masoero2022clustering} to better understand the performance of the clustering algorithm at a global scale.

\begin{figure}
         \includegraphics[width=1\textwidth]{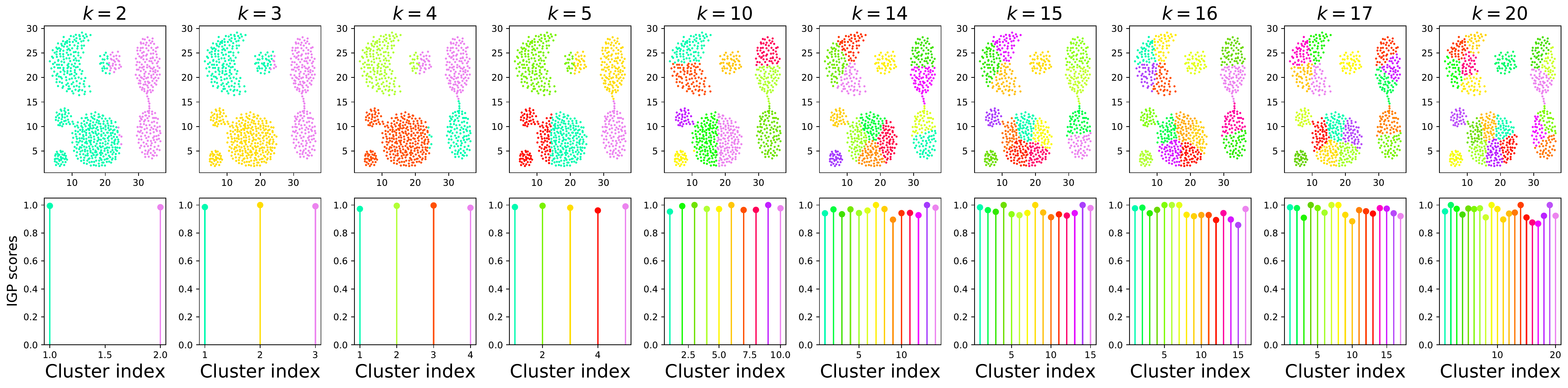} 
\caption{IGP computed for the ``aggregation'' dataset using $k$-means, for different values of $k$.}
\label{fig:IGP_aggregation}
\end{figure}

\begin{figure}
         \includegraphics[width=1\textwidth]{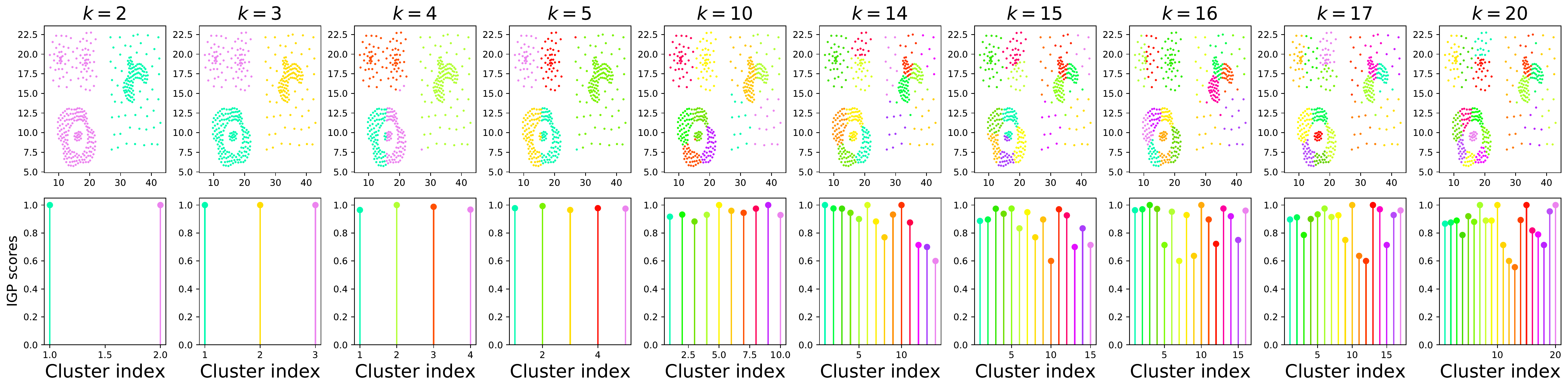} 
\caption{IGP computed for the ``compound'' dataset using $k$-means, for different values of $k$.}
\label{fig:IGP_compound}
\end{figure}

\begin{figure}
         \includegraphics[width=1\textwidth]{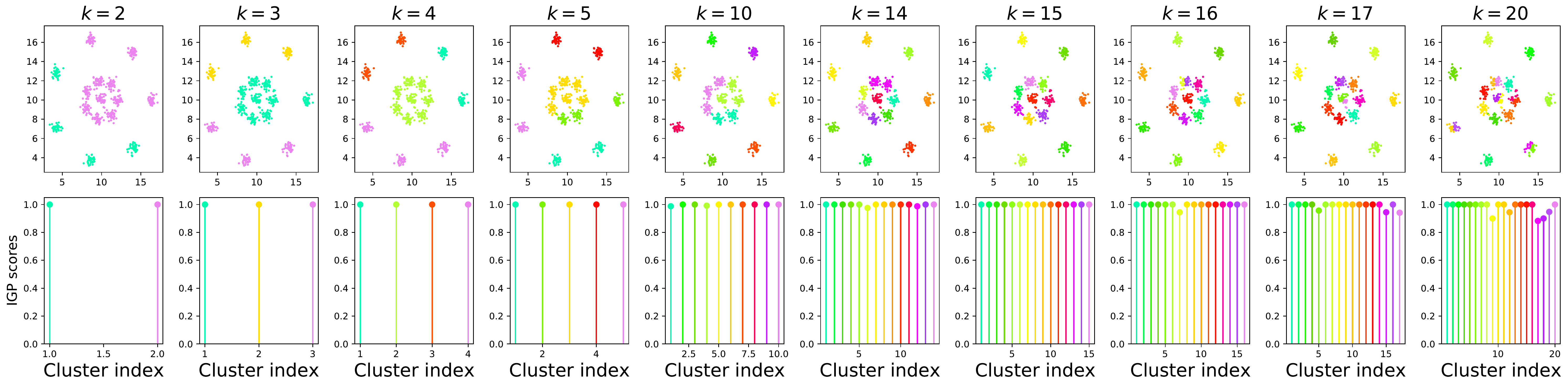} 
\caption{IGP computed for the ``r15'' dataset using $k$-means, for different values of $k$.}
\label{fig:IGP_r15}
\end{figure}

Next, we provide results for the IGP, first proposed in \citet{kapp2006clusters}, in \Cref{fig:IGP_aggregation,fig:IGP_compound,fig:IGP_r15}. In our experiments, we find that the IGP scores tend to provide an overconfident statement about the replicability of clustering algorithms, often providing scores close to $1$ (the highest possible value) also for spurious clusters (see, e.g., the case of $k=10$ for the ``aggregation'' dataset in \Cref{fig:IGP_aggregation}). We notice that both the prediction strength and the IGP do not require the specification of any additional parameter to assess replicability other than the choice of the clustering algorithm. In applications, it might be good practice to compute these scores on the original dataset as well as on perturbed versions of the original data, in a similar fashion to what is performed to obtain the D and R indices. Moreover, for the IGP, we suggest practitioners interested in using this metric to not only compute the metric using only each point's nearest neighbor, but also other points nearby (e.g., compute the score by checking whether all $j$-nearest neighbors belong to the same cluster, for different values of $j$).

%%%%%%%%%%%%%%%%%%%%%%%%
\subsection{Clustering replicability via tests of significance} \label{sec:tests_exp}
%%%%%%%%%%%%%%%%%%%%%%%%

\begin{figure}
         \includegraphics[width=1\textwidth]{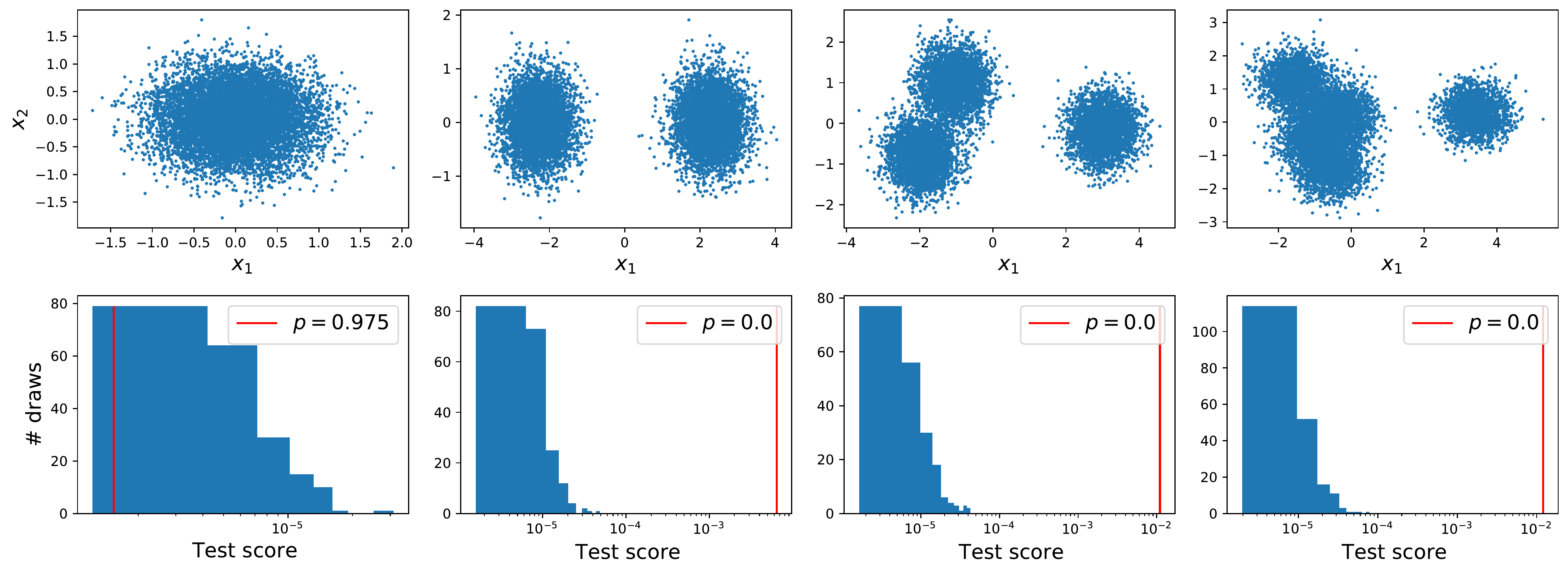} 
\caption{Statistical tests of significance to reject the null hypothesis $H_0$ that no clustering structure is present in the data.}
\label{fig:test_k1}
\end{figure}

\begin{figure}
         \includegraphics[width=1\textwidth]{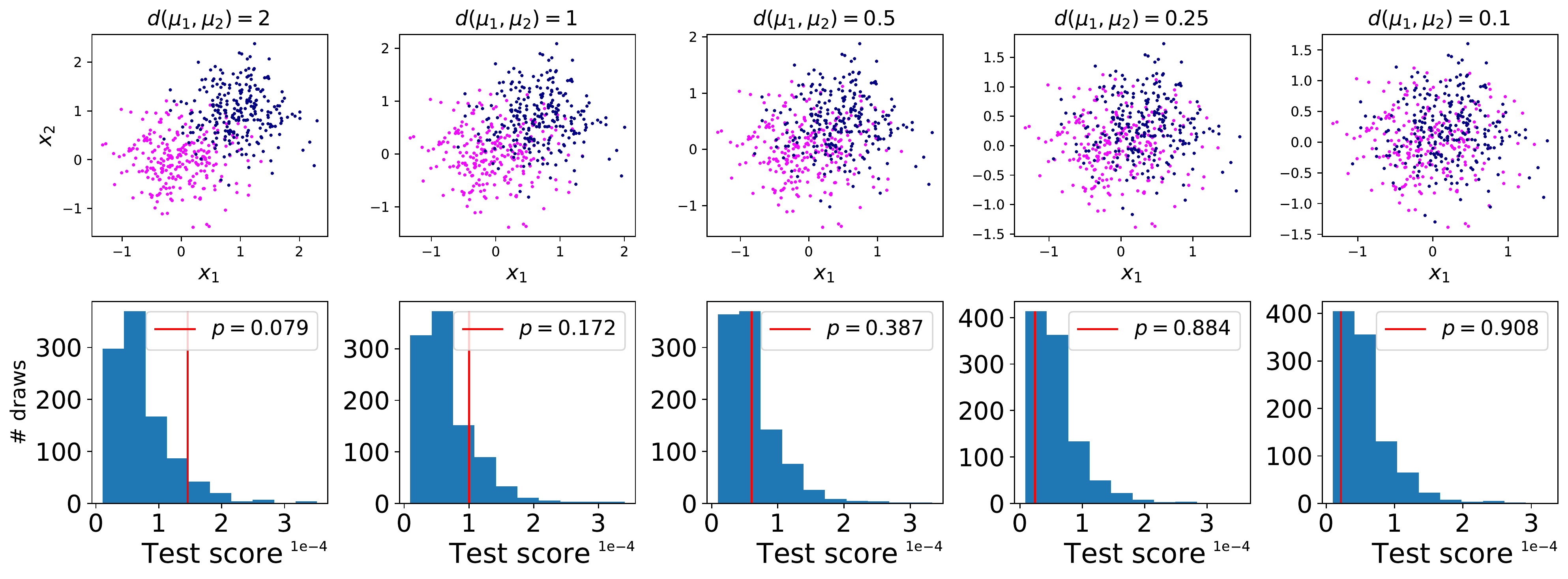} 
\caption{Statistical tests of significance to reject the null hypothesis $H_0$ that no clustering structure is present in the data.}
\label{fig:test_distance}
\end{figure}

Last, we discuss replicability metrics that rely on significance tests as discussed in Section 2.4. In particular, we follow the procedure proposed by \citet{mcshanerepro} and report results obtained on synthetic data in \Cref{fig:test_k1,fig:test_distance}.

First, in \Cref{fig:test_k1}, we draw $N=1000$ samples from Gaussian mixture models in $\R^2$ with $k$ components, for $k \in \{1,2,3,5\}$. The data is plotted in the first row of the figure. For each of these datasets, we test the null hypothesis $H_0$ of absence of clustering following the procedure discussed in Section 2.4, using a multivariate Gaussian distribution with mean given by the sample mean and covariance matrix given by the sample covariance matrix to draw from the null model. In our simulation, we let $B=200$ (i.e., we redraw $B=200$ datasets of size $N=1000$, and for each $b=1,\ldots,B$ of these datasets compute the score $s_b$ discussed in Section 2.4. Results (and corresponding $p$-value under the null) are reported in the second row of the figure. The test behaves as expected --- in particular, it successfully allows to reject the null hypothesis when a clear clustering structure is present in the data.

Next, we analyze the sensitivity of the test as a function of the separation between the clusters in the data. Towards this goal, we consider $N=1000$ datapoints drawn from balanced bi-variate Gaussian mixture models in $\R^2$,
\[
	F = \frac{1}{2} \mathcal{N}(\mu_1,I)+ \frac{1}{2}\mathcal{N}(\mu_2,I).
\]
We analyze how the test behaves as we decrease the distance between the means of the two components $\mu_1, \mu_2$. Results are reported along the columns of \Cref{fig:test_distance}. As expected, as we decrease the distance, and the clustering structure becomes less clear, the $p$-value associated with the test starts decreasing. For ease of visualization, we color points belonging to different clusters with different colors.

%% file: main.bbl
\begin{thebibliography}{51}
\providecommand{\natexlab}[1]{#1}
\providecommand{\url}[1]{\texttt{#1}}
\expandafter\ifx\csname urlstyle\endcsname\relax
  \providecommand{\doi}[1]{doi: #1}\else
  \providecommand{\doi}{doi: \begingroup \urlstyle{rm}\Url}\fi

\bibitem[Albatineh et~al.(2006)Albatineh, Niewiadomska-Bugaj, and
  Mihalko]{albatineh2006similarity}
A.~N. Albatineh, M.~Niewiadomska-Bugaj, and D.~Mihalko.
\newblock On similarity indices and correction for chance agreement.
\newblock \emph{Journal of Classification}, 23\penalty0 (2), 2006.

\bibitem[Alexe et~al.(2006)Alexe, Dalgin, Ramaswamy, DeLisi, and
  Bhanot]{alexe2006data}
G.~Alexe, G.~S. Dalgin, R.~Ramaswamy, C.~DeLisi, and G.~Bhanot.
\newblock Data perturbation independent diagnosis and validation of breast
  cancer subtypes using clustering and patterns.
\newblock \emph{Cancer Informatics}, 2, 2006.

\bibitem[Arrieta et~al.(2020)Arrieta, D{\'\i}az-Rodr{\'\i}guez, Del~Ser,
  Bennetot, Tabik, Barbado, Garc{\'\i}a, Gil-L{\'o}pez, Molina, and
  Benjamins]{arrieta2020explainable}
A.~B. Arrieta, N.~D{\'\i}az-Rodr{\'\i}guez, J.~Del~Ser, A.~Bennetot, S.~Tabik,
  A.~Barbado, S.~Garc{\'\i}a, S.~Gil-L{\'o}pez, D.~Molina, and R.~Benjamins.
\newblock Explainable artificial intelligence (xai): Concepts, taxonomies,
  opportunities and challenges toward responsible ai.
\newblock \emph{Information fusion}, 58:\penalty0 82--115, 2020.

\bibitem[Ben-David et~al.(2007)Ben-David, P{\'a}l, and Simon]{ben2007stability}
S.~Ben-David, D.~P{\'a}l, and H.~U. Simon.
\newblock Stability of $k$-means clustering.
\newblock In \emph{International Conference on Computational Learning Theory}.
  Springer, 2007.

\bibitem[Bernau et~al.(2014)Bernau, Riester, Boulesteix, Parmigiani,
  Huttenhower, Waldron, and Trippa]{bernau2014cross}
C.~Bernau, M.~Riester, A.-L. Boulesteix, G.~Parmigiani, C.~Huttenhower,
  L.~Waldron, and L.~Trippa.
\newblock Cross-study validation for the assessment of prediction algorithms.
\newblock \emph{Bioinformatics}, 30\penalty0 (12), 2014.

\bibitem[Bertoni and Valentini(2007)]{bertoni2007model}
A.~Bertoni and G.~Valentini.
\newblock Model order selection for bio-molecular data clustering.
\newblock \emph{BMC Bioinformatics}, 8\penalty0 (2), 2007.

\bibitem[Brock et~al.(2008)Brock, Pihur, Datta, and Datta]{brock2008clvalid}
G.~Brock, V.~Pihur, S.~Datta, and S.~Datta.
\newblock clvalid: An {R} package for cluster validation.
\newblock \emph{Journal of Statistical Software}, 25, 2008.

\bibitem[Bryan(2004)]{bryan2004problems}
J.~Bryan.
\newblock Problems in gene clustering based on gene expression data.
\newblock \emph{Journal of Multivariate Analysis}, 90\penalty0 (1), 2004.

\bibitem[Ester et~al.(1996)Ester, Kriegel, Sander, and Xu]{dbscan}
M.~Ester, H.-P. Kriegel, J.~Sander, and X.~Xu.
\newblock A density-based algorithm for discovering clusters in large spatial
  databases with noise.
\newblock \emph{Proceedings of the Second International Conference on Knowledge
  Discovery and Data Mining}, 1996.

\bibitem[Fang and Wang(2012)]{fang2012selection}
Y.~Fang and J.~Wang.
\newblock Selection of the number of clusters via the bootstrap method.
\newblock \emph{Computational Statistics \& Data Analysis}, 56\penalty0 (3),
  2012.

\bibitem[Fr{\"a}nti and Sieranoja(2018)]{ClusteringDatasets}
P.~Fr{\"a}nti and S.~Sieranoja.
\newblock K-means properties on six clustering benchmark datasets.
\newblock \emph{Applied Intelligence}, 48\penalty0 (12), 2018.

\bibitem[Fr\"anti et~al.(2006)Fr\"anti, Virmajoki, and Hautam\"aki]{DIMsets}
P.~Fr\"anti, O.~Virmajoki, and V.~Hautam\"aki.
\newblock Fast agglomerative clustering using a $k$-nearest neighbor graph.
\newblock \emph{IEEE Transactions on Pattern Analysis and Machine
  Intelligence}, 28\penalty0 (11), 2006.

\bibitem[Frey and Dueck(2007)]{frey2007clustering}
B.~J. Frey and D.~Dueck.
\newblock Clustering by passing messages between data points.
\newblock \emph{Science}, 315\penalty0 (5814), 2007.

\bibitem[Haibe-Kains et~al.(2012)Haibe-Kains, Desmedt, Loi, Culhane, Bontempi,
  Quackenbush, and Sotiriou]{haibe2012three}
B.~Haibe-Kains, C.~Desmedt, S.~Loi, A.~C. Culhane, G.~Bontempi, J.~Quackenbush,
  and C.~Sotiriou.
\newblock A three-gene model to robustly identify breast cancer molecular
  subtypes.
\newblock \emph{Journal of the National Cancer Institute}, 104\penalty0 (4),
  2012.

\bibitem[Hayes et~al.(2006)Hayes, Monti, Parmigiani, Gilks, Naoki,
  Bhattacharjee, Socinski, Perou, and Meyerson]{hayes2006gene}
D.~N. Hayes, S.~Monti, G.~Parmigiani, C.~B. Gilks, K.~Naoki, A.~Bhattacharjee,
  M.~A. Socinski, C.~Perou, and M.~Meyerson.
\newblock Gene expression profiling reveals reproducible human lung
  adenocarcinoma subtypes in multiple independent patient cohorts.
\newblock \emph{Journal of Clinical Oncology}, 24\penalty0 (31), 2006.

\bibitem[Hennig(2007)]{hennig2007cluster}
C.~Hennig.
\newblock Cluster-wise assessment of cluster stability.
\newblock \emph{Computational Statistics \& Data Analysis}, 52\penalty0 (1),
  2007.

\bibitem[Hennig(2015)]{hennig2015package}
C.~Hennig.
\newblock Package \emph{‘fpc’}.
\newblock \emph{R-project}, 91, 2015.

\bibitem[Hubert and Arabie(1985)]{hubert1985comparing}
L.~Hubert and P.~Arabie.
\newblock Comparing partitions.
\newblock \emph{Journal of Classification}, 2\penalty0 (1), 1985.

\bibitem[Jaskowiak et~al.(2014)Jaskowiak, Campello, and
  Costa]{jaskowiak2014selection}
P.~A. Jaskowiak, R.~J. Campello, and I.~G. Costa.
\newblock On the selection of appropriate distances for gene expression data
  clustering.
\newblock \emph{BMC Bioinformatics}, 15\penalty0 (2):\penalty0 S2, 2014.

\bibitem[Kapp and Tibshirani(2006)]{kapp2006clusters}
A.~V. Kapp and R.~Tibshirani.
\newblock Are clusters found in one dataset present in another dataset?
\newblock \emph{Biostatistics}, 8\penalty0 (1), 2006.

\bibitem[Lancaster and Seneta(1969)]{lancaster1969}
H.~O. Lancaster and E.~Seneta.
\newblock Chi-square distribution.
\newblock \emph{Encyclopedia of Biostatistics}, 2, 1969.

\bibitem[Lange et~al.(2004)Lange, Roth, Braun, and Buhmann]{lange2004stability}
T.~Lange, V.~Roth, M.~L. Braun, and J.~M. Buhmann.
\newblock Stability-based validation of clustering solutions.
\newblock \emph{Neural Computation}, 16\penalty0 (6), 2004.

\bibitem[Levenstien et~al.(2003)Levenstien, Yang, and
  Ott]{levenstien2003statistical}
M.~A. Levenstien, Y.~Yang, and J.~Ott.
\newblock Statistical significance for hierarchical clustering in genetic
  association and microarray expression studies.
\newblock \emph{BMC Bioinformatics}, 4\penalty0 (1), 2003.

\bibitem[Levine and Domany(2001)]{levine2001resampling}
E.~Levine and E.~Domany.
\newblock Resampling method for unsupervised estimation of cluster validity.
\newblock \emph{Neural Computation}, 13\penalty0 (11), 2001.

\bibitem[Lim and Yu(2016)]{lim2016estimation}
C.~Lim and B.~Yu.
\newblock Estimation stability with cross-validation (escv).
\newblock \emph{Journal of Computational and Graphical Statistics}, 25\penalty0
  (2):\penalty0 464--492, 2016.

\bibitem[Liu et~al.(2008)Liu, Hayes, Nobel, and Marron]{liu2008statistical}
Y.~Liu, D.~N. Hayes, A.~Nobel, and J.~S. Marron.
\newblock Statistical significance of clustering for high-dimension,
  low--sample size data.
\newblock \emph{Journal of the American Statistical Association}, 103\penalty0
  (483), 2008.

\bibitem[Lloyd(1982)]{lloyd1982least}
S.~Lloyd.
\newblock Least squares quantization in {PCM}.
\newblock \emph{IEEE Transactions on Information Theory}, 28\penalty0 (2),
  1982.

\bibitem[Maaten and Hinton(2008)]{maaten2008visualizing}
L.~v.~d. Maaten and G.~Hinton.
\newblock Visualizing data using t-{SNE}.
\newblock \emph{Journal of Machine Learning Research}, 9\penalty0 (Nov), 2008.

\bibitem[Masoero et~al.(2022{\natexlab{a}})Masoero, Thomas, Parmigiani,
  Tyekucheva, and Trippa]{masoero2022clustering}
L.~Masoero, E.~Thomas, G.~Parmigiani, S.~Tyekucheva, and L.~Trippa.
\newblock Cross-study replicability in cluster analysis.
\newblock \emph{Statistical Science}, 2022{\natexlab{a}}.

\bibitem[Masoero et~al.(2022{\natexlab{b}})Masoero, Thomas, Parmigiani,
  Tyekucheva, and Trippa]{masoero2022clusteringsupp}
L.~Masoero, E.~Thomas, G.~Parmigiani, S.~Tyekucheva, and L.~Trippa.
\newblock Supplementary matrial for ``cross-study replicability in cluster
  analysis''.
\newblock \emph{Statistical Science}, 2022{\natexlab{b}}.

\bibitem[McShane et~al.(2002)McShane, Radmacher, Freidlin, Yu, Li, and
  Simon]{mcshanerepro}
L.~M. McShane, M.~D. Radmacher, B.~Freidlin, R.~Yu, M.-C. Li, and R.~Simon.
\newblock Methods for assessing reproducibility of clustering patterns observed
  in analyses of microarray data.
\newblock \emph{Bioinformatics}, 18\penalty0 (11), 2002.

\bibitem[M{\"u}ller and Quintana(2010)]{muller2010random}
P.~M{\"u}ller and F.~Quintana.
\newblock Random partition models with regression on covariates.
\newblock \emph{Journal of Statistical Planning and Inference}, 140\penalty0
  (10):\penalty0 2801--2808, 2010.

\bibitem[Murdoch et~al.(2019)Murdoch, Singh, Kumbier, Abbasi-Asl, and
  Yu]{murdoch2019definitions}
W.~J. Murdoch, C.~Singh, K.~Kumbier, R.~Abbasi-Asl, and B.~Yu.
\newblock Definitions, methods, and applications in interpretable machine
  learning.
\newblock \emph{Proceedings of the National Academy of Sciences}, 116\penalty0
  (44):\penalty0 22071--22080, 2019.

\bibitem[National Academies~of Sciences and
  Medicine(2019)]{national2019reproducibility}
E.~National Academies~of Sciences and Medicine.
\newblock \emph{Reproducibility and Replicability in Science}.
\newblock The National Academies Press, Washington, DC, 2019.

\bibitem[Parker et~al.(2009)Parker, Mullins, Cheang, Leung, Voduc, Vickery,
  Davies, Fauron, He, Hu, Quackenbush, Stijleman, Palazzo, Marron, Nobel,
  Mardis, Nielsen, Ellis, Perou, and Bernard]{pam_50}
J.~S. Parker, M.~Mullins, M.~C. Cheang, S.~Leung, D.~Voduc, T.~Vickery,
  S.~Davies, C.~Fauron, X.~He, Z.~Hu, J.~F. Quackenbush, I.~J. Stijleman,
  J.~Palazzo, J.~Marron, A.~B. Nobel, E.~Mardis, T.~O. Nielsen, M.~J. Ellis,
  C.~M. Perou, and P.~S. Bernard.
\newblock Supervised risk predictor of breast cancer based on intrinsic
  subtypes.
\newblock \emph{Journal of Clinical Oncology}, 27\penalty0 (8), 2009.

\bibitem[Perou et~al.(2000)Perou, S{\o}rlie, Eisen, Van De~Rijn, Jeffrey, Rees,
  Pollack, Ross, Johnsen, and Akslen]{perou2000molecular}
C.~M. Perou, T.~S{\o}rlie, M.~B. Eisen, M.~Van De~Rijn, S.~S. Jeffrey, C.~A.
  Rees, J.~R. Pollack, D.~T. Ross, H.~Johnsen, and L.~A. Akslen.
\newblock Molecular portraits of human breast tumours.
\newblock \emph{Nature}, 406\penalty0 (6797), 2000.

\bibitem[Rand(1971)]{rand1971objective}
W.~M. Rand.
\newblock Objective criteria for the evaluation of clustering methods.
\newblock \emph{Journal of the American Statistical Association}, 66\penalty0
  (336), 1971.

\bibitem[Schroeder et~al.(2011{\natexlab{a}})Schroeder, Haibe-Kains, Culhane,
  Sotiriou, Bontempi, and Quackenbush]{mainz}
M.~Schroeder, B.~Haibe-Kains, A.~Culhane, C.~Sotiriou, G.~Bontempi, and
  J.~Quackenbush.
\newblock \emph{breastCancerMAINZ: Gene expression dataset published by Schmidt
  et al. [2008] (MAINZ).}, 2011{\natexlab{a}}.
\newblock R package version 1.16.0.

\bibitem[Schroeder et~al.(2011{\natexlab{b}})Schroeder, Haibe-Kains, Culhane,
  Sotiriou, Bontempi, and Quackenbush]{transbig}
M.~Schroeder, B.~Haibe-Kains, A.~Culhane, C.~Sotiriou, G.~Bontempi, and
  J.~Quackenbush.
\newblock \emph{breastCancerTRANSBIG: Gene expression dataset published by
  Desmedt et al. [2007] (TRANSBIG).}, 2011{\natexlab{b}}.
\newblock R package version 1.16.0.

\bibitem[Schroeder et~al.(2011{\natexlab{c}})Schroeder, Haibe-Kains, Culhane,
  Sotiriou, Bontempi, and Quackenbush]{vdx}
M.~Schroeder, B.~Haibe-Kains, A.~Culhane, C.~Sotiriou, G.~Bontempi, and
  J.~Quackenbush.
\newblock \emph{breastCancerVDX: Gene expression datasets published by Wang et
  al. [2005] and Minn et al. [2007] (VDX)}, 2011{\natexlab{c}}.
\newblock R package version 1.16.0.

\bibitem[Smolkin and Ghosh(2003)]{smolkin2003cluster}
M.~Smolkin and D.~Ghosh.
\newblock Cluster stability scores for microarray data in cancer studies.
\newblock \emph{BMC Bioinformatics}, 4\penalty0 (1), 2003.

\bibitem[Tibshirani and Walther(2005)]{tibshirani2005cluster}
R.~Tibshirani and G.~Walther.
\newblock Cluster validation by prediction strength.
\newblock \emph{Journal of Computational and Graphical Statistics}, 14\penalty0
  (3), 2005.

\bibitem[Trippa et~al.(2015)Trippa, Waldron, Huttenhower, and
  Parmigiani]{trippa2015bayesian}
L.~Trippa, L.~Waldron, C.~Huttenhower, and G.~Parmigiani.
\newblock Bayesian nonparametric cross-study validation of prediction methods.
\newblock \emph{The Annals of Applied Statistics}, 9\penalty0 (1), 2015.

\bibitem[Vinh et~al.(2009)Vinh, Epps, and Bailey]{vinh2009information}
N.~X. Vinh, J.~Epps, and J.~Bailey.
\newblock Information theoretic measures for clusterings comparison: is a
  correction for chance necessary?
\newblock In \emph{Proceedings of the 26th Annual International Conference on
  Machine Learning}. ACM, 2009.

\bibitem[Vinh et~al.(2010)Vinh, Epps, and Bailey]{vinh2010information}
N.~X. Vinh, J.~Epps, and J.~Bailey.
\newblock Information theoretic measures for clusterings comparison: Variants,
  properties, normalization and correction for chance.
\newblock \emph{Journal of Machine Learning Research}, 11\penalty0 (Oct), 2010.

\bibitem[Von~Luxburg(2010)]{von2010clustering}
U.~Von~Luxburg.
\newblock Clustering stability: an overview.
\newblock \emph{Foundations and Trends in Machine Learning}, 2\penalty0 (3),
  2010.

\bibitem[Wade and Ghahramani(2018)]{wade2018bayesian}
S.~Wade and Z.~Ghahramani.
\newblock Bayesian cluster analysis: Point estimation and credible balls (with
  discussion).
\newblock \emph{Bayesian Analysis}, 13\penalty0 (2):\penalty0 559--626, 2018.

\bibitem[Waks and Winer(2019)]{waks2019breast}
A.~G. Waks and E.~P. Winer.
\newblock Breast cancer treatment: A review.
\newblock \emph{JAMA}, 321\penalty0 (3), 2019.

\bibitem[Ward~Jr(1963)]{ward1963hierarchical}
J.~H. Ward~Jr.
\newblock Hierarchical grouping to optimize an objective function.
\newblock \emph{Journal of the American Statistical Association}, 58\penalty0
  (301), 1963.

\bibitem[Yu(2013)]{yu2013stability}
B.~Yu.
\newblock Stability.
\newblock \emph{Bernoulli}, 19\penalty0 (4), 2013.

\bibitem[Zhang et~al.(1996)Zhang, Ramakrishnan, and Livny]{zhang1996birch}
T.~Zhang, R.~Ramakrishnan, and M.~Livny.
\newblock Birch: an efficient data clustering method for very large databases.
\newblock In \emph{ACM Sigmod Record}. ACM, 1996.

\end{thebibliography}
